\documentclass[usenatbib,useAMS,usedcolumn]{mn2e}
\usepackage{times}

\usepackage{epsfig}

\newcommand{\arcm}{\hbox{$^\prime$}}

\newcommand{\degree}{\hbox{$^\circ$}}

\newcommand{\chandra}{\emph{Chandra}}
\newcommand{\xmm}{\emph{XMM-Newton}}
\newcommand{\xmms}{\emph{XMM}}
\newcommand{\asca}{\emph{ASCA}}

\newcommand{\arcs}{\mbox{\arcm\arcm}}

\newcommand{\Zsol}{\ensuremath{~Z_{\odot}}}

\newcommand{\Msol}{\ensuremath{~M_{\odot}}}
\newcommand{\LB}{\ensuremath{L_{\mathrm{B}}}}
\newcommand{\LBsol}{\ensuremath{L_{B\odot}}}

\newcommand{\s}{\ensuremath{\mbox{~s}}}
\newcommand{\ps}{\ensuremath{\s^{-1}}}
\newcommand{\cm}{\ensuremath{\mbox{~cm}}}
\newcommand{\pcmsq}{\ensuremath{\cm^{-2}}}
\newcommand{\pcmcu}{\ensuremath{\cm^{-3}}}
\newcommand{\km}{\ensuremath{\mbox{~km}}}

\newcommand{\erg}{\ensuremath{\mbox{~erg}}}
\newcommand{\ergps}{\ensuremath{\erg \ps}}

\newcommand{\ergpcmcu}{\ensuremath{\erg \pcmcu}}
\newcommand{\kmps}{\ensuremath{\km \ps}}
\newcommand{\mJy}{\ensuremath{\mbox{~mJy}}}
\newcommand{\bm}{\ensuremath{\mbox{~b}}}
\newcommand{\pb}{\ensuremath{\bm^{-1}}}
\newcommand{\mJypb}{\ensuremath{\mJy \pb}}

\newcommand{\Ho}{\ensuremath{\mathrm{H_0}}}

\newcommand{\Dtf}{\ensuremath{D_{\mathrm{25}}}}

\begin{document}

\title[
A deep \textit{Chandra} observation of AWM~4
]{
A deep \textit{Chandra} observation of the poor cluster AWM~4 -- I. Properties of the central
radio galaxy and its effects on the intracluster medium
}

\author[E. O'Sullivan et al.]
{Ewan O'Sullivan$^{1,2}$\thanks{E-mail: ejos@star.sr.bham.ac.uk (EO'S)}, Simona Giacintucci$^{1,3}$, Laurence P. David$^{1}$, Jan M. Vrtilek$^{1}$ \newauthor and Somak Raychaudhury$^{2}$\\
$^{1}$ Harvard-Smithsonian Center for Astrophysics, 60 Garden Street,
Cambridge, MA 02138, USA \\ 
$^{2}$ School of Physics and Astronomy, University of Birmingham, Edgbaston, B15 2TT, UK \\
$^{3}$ INAF -- Instituto di Radioastronomia, via Gobetti 101, 40129 Bologna, Italy
}

\date{Accepted ....; Received ....; in original form 22 March 2010}

\pagerange{\pageref{firstpage}--\pageref{lastpage}} \pubyear{2010}

\label{firstpage}

\maketitle

\begin{abstract}
  Using observations from the \textit{Chandra} X--ray Observatory and Giant
  Metrewave Radio Telescope, we examine the interaction between the
  intracluster medium and central radio source in the poor cluster AWM~4.
  In the \textit{Chandra} observation a small cool core or galactic corona
  is resolved coincident with the radio core. This corona is capable of
  fuelling the active nucleus, but must be inefficiently heated by jet
  interactions or conduction, possibly precluding a feedback relationship
  between the radio source and cluster.  A lack of clearly detected X--ray
  cavities suggests that the radio lobes are only partially filled by
  relativistic plasma. We estimate a filling factor of $\phi$=0.21
  (3$\sigma$ upper limit $\phi<0.42$) for the better constrained east lobe.
  We consider the particle population in the jets and lobes, and find that
  the standard equipartition assumptions predict pressures and ages which
  agree poorly with X--ray estimates. Including an electron population
  extending to low Lorentz factors either reduces ($\gamma_{min}=100$) or
  removes ($\gamma_{min}=10$) the pressure imbalance between the lobes and
  their environment.  Pressure balance can also be achieved by entrainment
  of thermal gas, probably in the first few kiloparsecs of the radio jets.
  We estimate the mechanical power output of the radio galaxy, and find it
  to be marginally capable of balancing radiative cooling.
\end{abstract}


\section{Introduction}
\label{sec:intro}

X--ray observations of clusters and groups of galaxies over the last decade
have led to a significant revision of our models of the intergalactic
medium in these systems. The \chandra\ and \xmm\ observatories have
provided strong evidence that despite cooling times being significantly
shorter than the Hubble time \citep[e.g.,][]{Sandersonetal06}, relatively
little gas actually cools below $\sim$1~keV
\citep{Petersonetal03,Kaastraetal04}.  It is now widely accepted that
excessive cooling is in many systems prevented by a feedback mechanism in
which the AGN of the central dominant galaxy, fuelled by cooling
intra--cluster gas, can reheat the gas through a variety of mechanisms
\citep[e.g.,][and references therein]{PetersonFabian06,McNamaraNulsen07}.

X--ray and radio images provide numerous examples of interactions between
AGN and the surrounding intra--cluster medium (ICM). Deep multiwavelength
observations of the brightest nearby clusters have revealed complex
structures associated with the radio jets and lobes, including shocks,
sound waves, individual and linked chains of cavities, uplifted material
and cooling filaments
\citep[e.g.,][]{Fabianetal05,Fabianetal06,Formanetal07,Wiseetal07,Blantonetal09}.
Much of this work has concentrated on the cavities in the ICM which radio
lobes produce as they inflate. The enthalpy of the cavities can be used as
a measure of the mechanical power output of the radio jets, and has been
shown to be sufficient to prevent or greatly reduce cooling in many systems
\citep{Birzanetal04}, provided the energy can be transfered into the
intracluster medium and distributed relatively isotropically.

The disturbed structures produced by AGN jet/ICM interactions are
relatively short--lived, and increasingly difficult to detect as they age.
The radio lobes which inflate cavities fade rapidly once the AGN outburst
ceases, and the X--ray cavities, which are detected by contrast with their
surroundings, become less visible once they move beyond the dense group or
cluster core. It is therefore considerably more difficult to study older
AGN outbursts. However, since the mechanism by which cavities heat their
surroundings is still a matter of debate, it is desirable to observe
systems with as wide a range of ages as possible, so as to understand
clearly the interaction between the radio lobes and their environment.

Most observations of radio galaxies to date have been based on observations
at frequencies $>$1~GHz. Radio lobes may be studied over a wider range of
timescales by observations at lower radio frequencies, which probe lower
energy electrons less affected by spectral aging. Deep X--ray imaging is
needed to complement such observations, and in this paper we discuss one
example where this combination is available, the poor cluster AWM~4.

A previous \xmm\ observation found AWM~4 to be approximately isothermal to
a radius of $\sim$150~kpc, with no evidence of a central cool core
\citep[hereafter referred to as OS05]{OSullivanetal05_special}. Comparison
with MKW~4, a cluster of similar temperature and galaxy population, but
which hosts a large cool core and lacks a central radio source
\citep{OSullivanetal03}, leads to the suggestion that AWM~4 had been strongly
heated by its central radio galaxy, 4C+24.36. However, examination of the
\xmm\ data showed no evidence of cavities or shocks, and no high resolution
images showing the lobes were available in the literature or radio
archives. The existing VLA 1.4~GHz data were interpreted as evidence
against the presence of lobes of sufficient volume to be responsible for
reheating a large cool core \citep{Gastaldelloetal08}. The ICM and galaxy
distribution both appear relaxed with no significant substructure, with a
strong concentration of early--type galaxies toward the core
\citep{KoranyiGeller02}. A cluster merger, which could also have heated the
ICM, therefore appears unlikely. The central elliptical, NGC~6051, shows no
signs of recent interactions \citep{Schombert87} and is considerably more
luminous than its neighbours, with a difference in magnitude above the
second--ranked galaxy of $M_{12}$=1.6 (SDSS $g$-band).

An analysis of deep GMRT radio observations at 235, 327 and 610~MHz was
presented in \citep[hereafter referred to as
GVM08]{Giacintuccietal08_special}. These data provided much new information
about 4C+24.36, revealing radio emission from the jets and lobes extending
$\sim$75~kpc from the AGN. The source was shown to be a wide--angle--tail
radio galaxy with inner jets oriented close to the plane of the sky,
probably moving southward with a velocity of $\la$120\kmps.  From modelling
of the progressive steepening of the spectral index $\alpha$ (defined as
$S\propto\nu^{-\alpha}$ where $S$ is flux and $\nu$ frequency) along the
jets the radiative age of the electron population was estimated as
160-170~Myr.

In this paper, we use a new \chandra\ ACIS-S observation of AWM~4, in
combination with the GMRT and archival VLA observations, to study the
structure of the ICM and the interaction of the AGN, radio jets and lobes
with the surrounding hot gas. The general properties of the cluster are
summarised in Table~\ref{tab:intro}, along with the position, distances and
angular scale of the system. Throughout the paper we assume \Ho=70,
$\Omega_M=0.3$, and $\Omega_{\Lambda}=0.7$. Uncertainties are generally
quoted at the 1$\sigma$ level, except in the case of X--ray spectral
fitting, where 90 percent uncertainties were estimated.
Section~\ref{sec:obs} describes the observation and data reduction, and
Sections~\ref{sec:img} and \ref{sec:spec} our imaging and spectral
analysis. In Section~\ref{sec:coregas} we examine the properties of the gas
in the core of NGC~6051, immediately surrounding the AGN, and in
Section~\ref{sec:PB} we discuss the interaction between the radio jets and
ICM, and place limits on the timescale of the outburst and the particle
content of the radio lobes. We discuss our results in
Section~\ref{sec:discuss} and list our conclusions in
Section~\ref{sec:con}.

\begin{table}
\caption{\label{tab:intro} General properties of the AWM~4 system}
\begin{tabular}{llc}
\hline
AWM~4    & z & 0.0318 \\
         & D$_A$ (Mpc) & 130.9 \\
         & D$_L$ (Mpc) & 139.3 \\
         & angular scale (kpc/\arcs) & 0.63 \\
         & $\sigma_v$ (\kmps) & 400 \\
NGC~6051 & RA$_{\rm J2000}$ (h m s) & 16 04 56.8 \\
         & DEC$_{\rm J2000}$ (\degree\ \arcm\ \arcs) & +23 55 56 \\
         & \LB\ (\LBsol) & 6.94$\times10^{10}$ \\
4C+24.36 & S$_{1.4 GHz, NVSS}$ (mJy) & 608 \\
         & S$_{235 MHz}$ (mJy) & 2750 \\
         & log P$_{1.4 GHz}$ (W Hz$^{-1}$) & 24.14 \\
\hline
\end{tabular}
\end{table}

\section{Observations and Data Reduction}
\label{sec:obs}
AWM~4 was observed by the \chandra\ ACIS instrument during Cycle~9 on 2008
May 18-19 
(ObsID 9423) for $\sim$80~ks. A summary of the \chandra\ mission and
instrumentation can be found in \citet{Weisskopfetal02}. The S3 CCD was
placed at the focus of the telescope and the instrument operated in very
faint mode, to take advantage of the superior cosmic ray rejection. We have
reduced the data from the pointing using CIAO 4.1.2 and CALDB 4.1.2
following techniques similar to those described in \citet{OSullivanetal07}
and the \chandra\ analysis
threads\footnote{http://asc.harvard.edu/ciao/threads/index.html}.  The
level 1 events files were reprocessed using the very faint mode filtering,
bad pixels and events with \asca\ grades 1, 5 and 7 were removed, and the
cosmic ray afterglow correction was applied. The data were corrected to the
appropriate gain map, the standard time-dependent gain and charge-transfer
inefficiency (CTI) corrections were made, and a background light curve was
produced.  The observation did not suffer from significant background
flaring, and the final cleaned exposure time was 74.5~ks. While data from
the entire detector were examined, for the purposes of this study we
generally only make use of the S3 CCD, as the radio source and the cluster
core fall on that chip.

Identification of point sources on S3 was performed using the \textsc{ciao}
task \textsc{wavdetect}, with a detection threshold of 10$^{-6}$, chosen to
ensure that the task detects $\leq$1 false source in the field, working
from a 0.3-7.0 keV image and exposure map.  Source ellipses were generated
with axes of length 4 times the standard deviation of each source
distribution. These were then used to exclude sources from most spectral
fits. A source was detected coincident with the peak of the diffuse X-ray
emission; this was considered a potentially false detection and ignored,
though we did later test for the presence of a central X-ray point source.

Spectra were extracted using the \textsc{specextract} task. Spectral
fitting was performed in XSPEC 11.3.2ag. Abundances were measured relative
to the abundance ratios of \citet{GrevesseSauval98}. A galactic hydrogen
column of 0.05$\times10^{22}$\pcmsq\ and a redshift of 0.0318 were assumed
in all fits (except those whose purpose was to test the effect of varying
hydrogen column), and 90\% errors are reported for all fitted values.
Spectra were grouped to 20 counts per bin, and counts at energies above 7
keV were ignored during fitting.

Background spectra were drawn from the standard set of CTI-corrected ACIS
blank sky background events files in the \chandra\ CALDB. The exposure time
of each background events file was altered to produce the same 9.5-12.0 keV
count rate as that in the target observation.  The same very faint mode
background screening was applied to the background data sets. Comparison of
source and background spectra suggested a slight excess of soft emission in
the background datasets, mainly below 0.5 keV. This is not unexpected, as
the soft X-ray background arises largely from hot gas in the galaxy, and
from coronal emission associated with solar wind interactions, and thus is
both spatially and temporally variable
\citep[e.g.,][]{KuntzSnowden00,Snowdenetal04}. There are also indications
that the spectral shape of the background has changed since the creation of
the blank-sky background files (c.f. the ACIS background
cookbook\footnote{http://asc.harvard.edu/contrib/maxim/acisbg/COOKBOOK}),
which could contribute to the disagreement at low energies. Experimenting
with fitting spectra in different energy bands, we found that ignoring
energies below 0.7 keV produces results which are consistent with those
derived from \xmm\ in OS05. If the absorbing hydrogen column is allowed to
vary in these fits it still tends to produce best fit values in excess of
the galactic value, but typically consistent with it at the 90\%
uncertainty level, and with the hydrogen column derived from the \xmm\
fits.  Temperature, abundance and normalisation of the models are not
significantly affected if the hydrogen column is allowed to vary, and our
results are therefore independent of its value.

\section{Imaging Analysis}
\label{sec:img}
\begin{figure}
\centerline{\includegraphics[width=\columnwidth,]{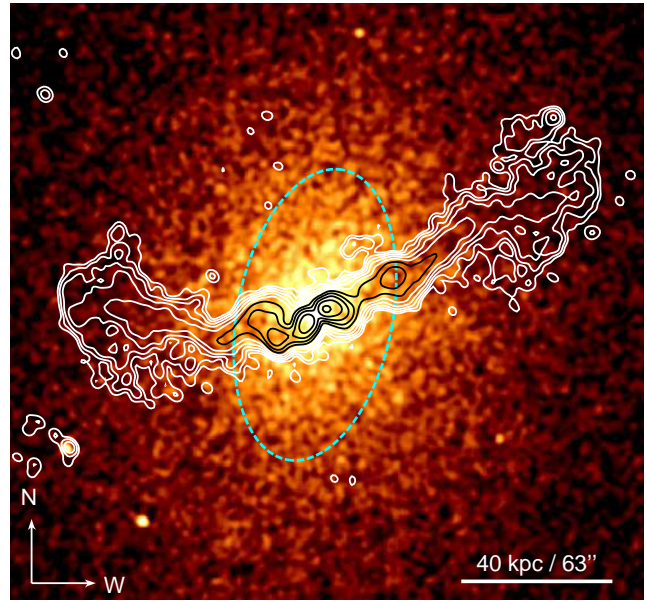}}
\caption{\label{fig:im} 0.3-2 keV exposure--corrected \chandra\ image of
  AWM~4, smoothed with a Gaussian of width 5 pixels. The dashed blue line
  marks the $R$-band \Dtf\ ellipse for the cD galaxy, NGC~6051. Contours
  (coloured black or white for clarity) represent the GMRT 610~MHz radio
  intensity, taken from GVM08, and are spaced by factors of 2, starting at
  0.2 \mJypb. The restoring beam was
  $5^{\prime\prime}\times4^{\prime\prime}$.}
\end{figure}

An initial examination of exposure--corrected images of AWM~4 agrees with
the general conclusions of the \xmm\ analysis of OS05. The cluster appears
to be relaxed, with a fairly smooth elliptical surface brightness
distribution roughly centred on NGC~6051. Figure~\ref{fig:im} shows a
soft-band image of the cluster with the \Dtf\ ellipse of NGC~6051 and GMRT
610~MHz contours overlaid. There are no surface brightness discontinuities
or ``fronts'' which would indicate the presence of large-scale gas motions
or shocks in the cluster halo.  However, a number of features which were
not observed in the \xmms\ images are visible in the \chandra\ data. The
most obvious of them is a bright central peak, closely aligned with the
central cD and the radio core detected at 4.9~GHz. This is only a few
pixels in diameter, and so could not have been resolved by \xmms. We will
discuss this in more detail in Section~\ref{sec:core}.

\begin{figure}
\centerline{\includegraphics[width=\columnwidth,]{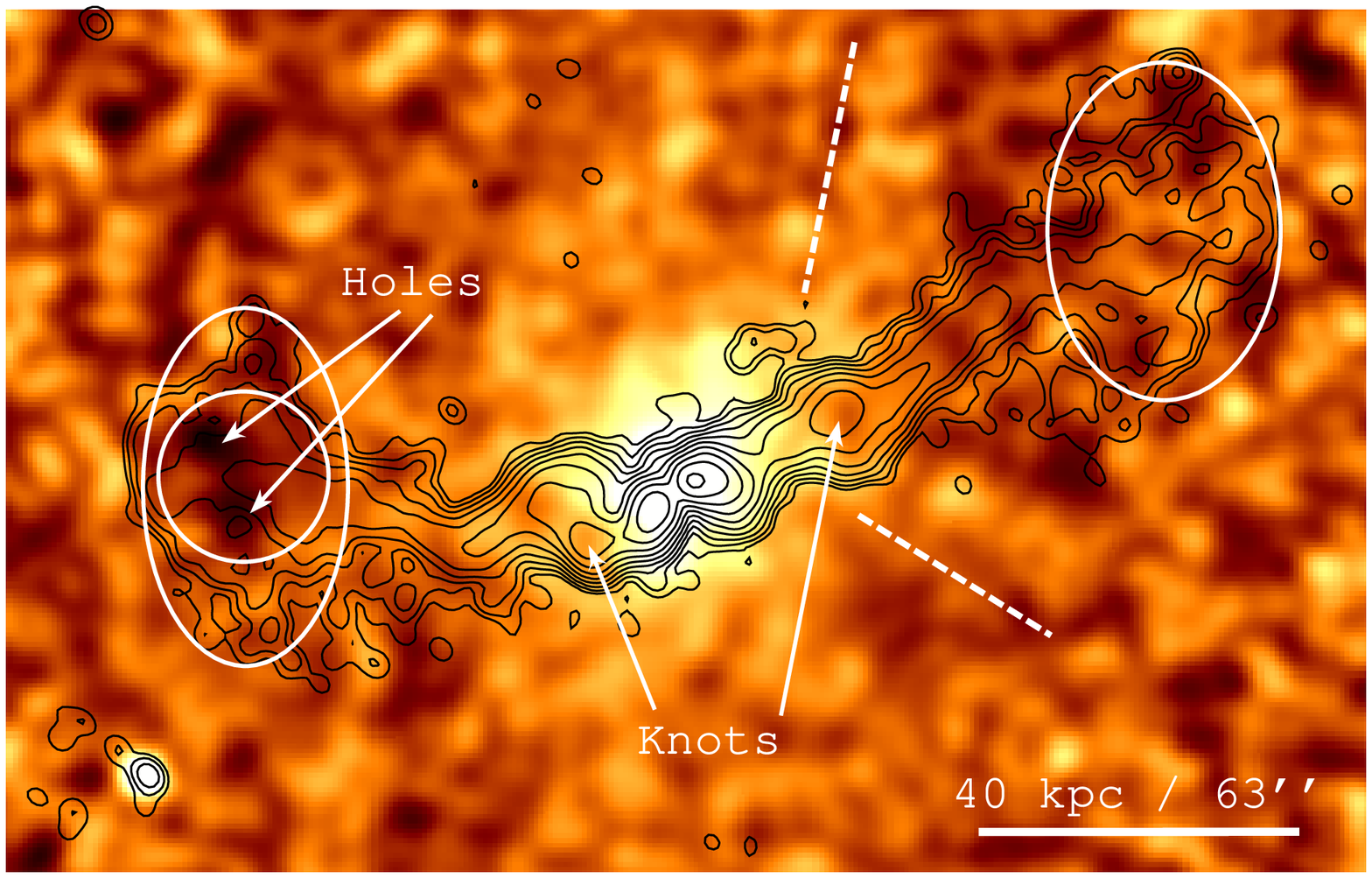}}
\centerline{\includegraphics[width=\columnwidth,]{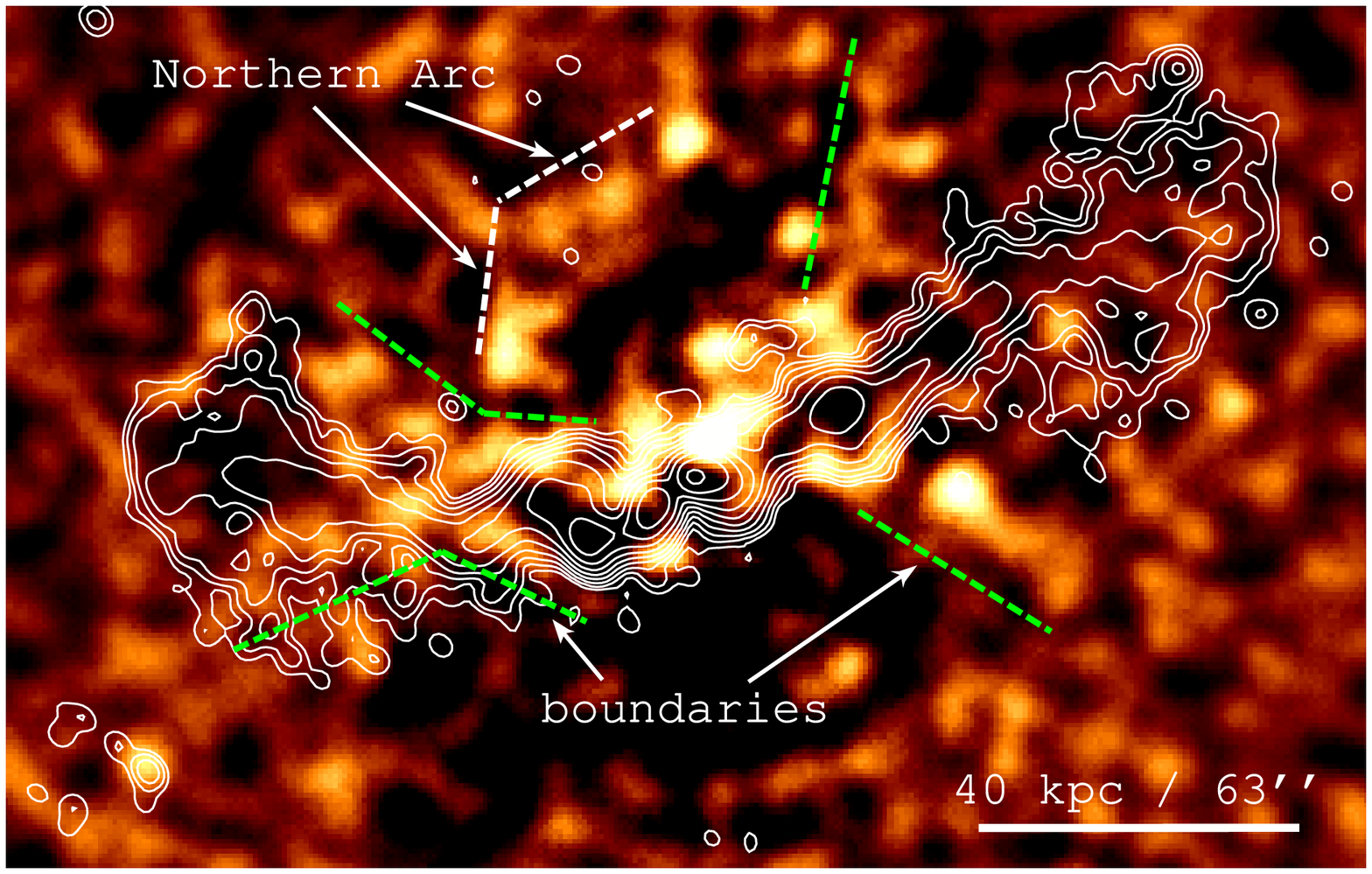}}
\caption{\label{fig:lobes} \textit{Upper panel}: \chandra\ 1-3~keV unsharp
  mask image, based on images binned by factor 2 and smoothed with 3 and 30
  pixel Gaussians. \textit{Lower panel}: 0.7-3~keV residual image, after
  subtraction of best-fitting surface brightness model, smoothed with a 7
  pixel Gaussian. 610~MHz radio contours are overlaid on both images (see
  Fig.~\ref{fig:im} for details). Regions discussed in the text are marked;
  ellipses and circles mark regions used for significance calculations,
  while dashed lines indicate possible structures in the X--ray emission.}
\end{figure}

Weak features are visible along the jets, particularly in the east lobe and
the knots $\sim$20~kpc from the radio core.  To examine these, we created
unsharp masked images smoothed at a range of scales.
Figure~\ref{fig:lobes} shows one example. The strongest feature in the
field is a surface brightness deficit in the east radio lobe. The deficit
is not well--correlated with the radio structure, instead appearing as two
small ``holes'' in the X--ray emission toward the end of the lobe. No
surface brightness structures are seen in the western lobe, though there is
some indication of a broad opening or bay in the surface brightness on the
west side of the cluster, into which the jet flows.

Using the \textsc{ciao sherpa} application we performed a two-dimensional
surface brightness fit to the S3 0.7-3 keV image and subtracted the model
image to search for residual features. Testing showed that PSF convolution
did not significantly affect the fit, and we therefore used an unconvolved
model. The background contribution was estimated from the scaled blank-sky
dataset, and we corrected for vignetting and other effects using a
monoenergetic 1.05~keV exposure map, the energy chosen to match the mean
photon energy of the data. A reasonably accurate fit was produced with two
$\beta$-models, an extended elliptical component describing the large-scale
emission and a compact component with negligible ellipticity describing the
core surface brightness peak. The best fit parameters are given in
Table~\ref{tab:SB} and the residual image is shown in
Figure~\ref{fig:lobes}.

\begin{table}
\caption{\label{tab:SB} Parameters of 2-d surface brightness model used to
  create residual map}
\begin{center}
\begin{tabular}{lll}
Component & Parameter & Value (1$\sigma$ error) \\
\hline\\[-3mm]
1 & $r_{core}$ (\arcs) & 0.14$^{+0.11}_{-0.14}$ \\
  & $\beta$            & 0.433$^{+0.018}_{-0.023}$ \\
2 & $r_{core}$ (\arcs) & 39.90$^{+1.42}_{-1.18}$ \\
  & $\beta$            & 0.403$^\pm0.002$ \\
  & ellip.             & 0.231$\pm0.007$ \\
  & p.a. (\degree)     & 80.2$^\pm0.8$ \\
\hline
\end{tabular}
\end{center}
\medskip
Position angle (p.a.) represents the angle of the major axis north of due west.
\end{table}

Unfortunately for our purposes, the strongest features in the resulting
residual map are associated with a general inequality in surface brightness
between the north and south quadrants of the cluster. This is visible in
the residual image as a strong deficit in surface brightness to the south
of the radio jet. AWM~4 is imperfectly described by a simple elliptical
model, owing to the presence of additional surface brightness to the north
of the core. This suggests that NGC~6051 is located somewhat to the south
of the centroid of the large-scale X--ray halo. GVM08 determined that
NGC~6051 is probably moving south, and this may explain the offset between
the galaxy and ICM centroid. There is also an apparent arc of brighter
emission to the north of the core, whose origin is unclear.

The residual image does reveal some indications of structures associated
with the jets and lobes, most notably X--ray bright regions along the
boundaries of the jets and possibly the inner part of the eastern lobe. The
broad opening or ``bay'' in the surface brightness on the western side of
the cluster also produces some features, though the opening angle is
considerably wider than that of the jet and lobe.  All of these features
are relatively weak and identified in part because of apparent correlations
with the jets. 

\begin{figure}
\centerline{\includegraphics[width=\columnwidth,bb=20 200 570 750]{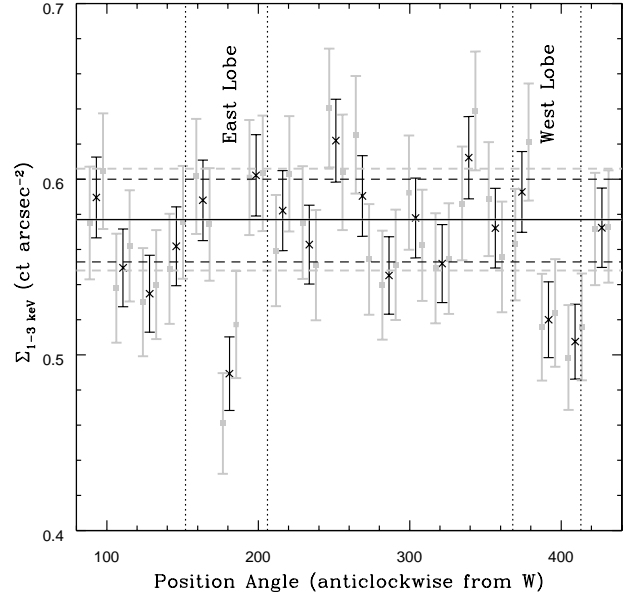}}
\caption{\label{fig:azimuth} 1-3~keV Azimuthal surface brightness profile
  around AWM~4 at the radius of the eastern lobe. Black points show
  18\degree\ sectors, grey points 9\degree\ sectors. The solid line marks
  the mean surface brightness excluding the 3 lowest black (6 lowest grey)
  points, dashed lines show the 1$\sigma$ uncertainties on the mean using
  each set. Dotted lines mark the angle within which the radio lobes are
  detected at 235~MHz.}
\end{figure}

To test the statistical significance of the features, we compared the
structures with regions at similar radii, using the surface brightness
models to define approximate isophotal ellipses.  An elliptical annulus of
width 40\arcs\ covers the surface brightness deficits in the eastern lobe
and the western bay, running through the base of the western lobe and the
centre of the eastern lobe. We measured the 1-3~keV exposure corrected
surface brightness (with point sources removed) in an azimuthal profile
around the ellipse, using sections with angular width 9\degree or
18\degree.  Figure~\ref{fig:azimuth} shows the results.

The two strongest negative features in the azimuthal profile occur within
the boundaries of the radio lobes, and correlate with the features seen in
the images. The deficit in the east lobe is stronger but smaller
(corresponding to the ``holes'') while the western deficit is broader but
less deep, and offset to the north side of the lobe.

We estimate the significance of the deficits in the lobes by calculating
the mean number of counts in the regions of the azimuthal profiles.
Excluding the regions within the lobes (as marked on
Figure~\ref{fig:azimuth}) the mean of the 9\degree\ regions (or 18\degree),
exposure corrected and normalised to a fixed area of 4600 (2300) pixels, is
640.9$\pm$25.3 (320.5$\pm$17.9) counts. The lowest value in the east lobe
is 547.1 (257.7) counts, giving a 3.7$\sigma$ (3.5$\sigma$) significance.
This suggests that the ``holes'' in the east lobe represent a significant
surface brightness depression, probably indicating the presence of a cavity
in the ICM.

The lowest points in the western lobe are 2.9$\sigma$ (2.3$\sigma$)
significant, and given the poorer alignment with the radio lobe it seems
possible that they represent a more general deficit on this side of the
cluster, rather than a cavity associated with the western lobe. This is
supported by a simple analysis of the X--ray surface brightness in an
ellipse chosen to match the lobe as a whole, compared with its surroundings.
This shows a deficit only at 1$\sigma$ significance.  Similar analysis for
the east and west jet knots yields low significances. The highest, for the
west knot, is 2.5$\sigma$. Deeper observations would be needed to determine
the nature of these apparent structures.

The lack of clear, significant structures in the cluster is in itself
interesting. If the radio lobes have excavated cavities in the ICM, we
would expect a surface brightness deficit on their line of sight.  We can
estimate the volume of the lobes from their projected size. We assume them
to be oblate ellipsoids with major/minor axes 21$\times$14.5~kpc (West) and
22.3$\times$12.8~kpc (East) at radii of 68.8 and 64.7~kpc respectively.
From the surface brightness models we expect to see 1120 counts (1-3~keV)
along the line of sight of the west lobe and 1270 from the east lobe if
there are no cavities. Based on the density profile of the ICM (see
Section~\ref{sec:spec}), we estimate that we would expect deficits of 184
counts and 251 counts for the west and east lobes respectively if the lobes
are empty of ICM plasma. We would therefore expect to detect cavities at
reasonable significance (4-5$\sigma$). The fact that we find a significant
decrement only in a small region of the east lobe raises questions about
the lobe contents.

\subsection{The cluster core and AGN}
\label{sec:core}
\begin{figure}
\centerline{\includegraphics[width=\columnwidth,]{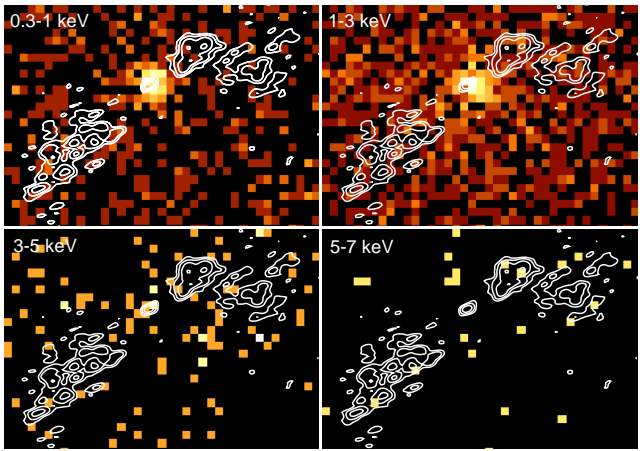}}
\caption{\label{fig:4band} Unbinned \chandra\ images of the core of AWM~4
  in four energy bands, with VLA 4.9~GHz contours overlaid. The energy
  bands are (upper left) 0.3-1 keV, (upper right) 1-3 keV, (lower left) 3-5
  keV and (lower right) 5-7 keV. The contour levels are spaced by factors
  of two starting at 0.15 \mJypb\ and are taken from an image with
  half-power beam width (HPBW) 0.7\arcs$\times$0.4\arcs. The images are
  $\sim$12~kpc or $\sim$19\arcs\ wide.}
\end{figure}

The observed central surface brightness peak could arise from several
causes, including X-ray emission from the AGN, the presence of a
small-scale cooling region or galactic corona, and emission associated with
the stellar population of NGC~6051. Surface brightness fitting to the 0.7-3
keV image confirms that the emission is extended and is poorly described by
a point source model, indicating that an AGN cannot be the only
source present. Figure~\ref{fig:4band} shows images in four energy bands,
0.3-1 keV, 1-3 keV, 3-5 keV and 5-7 keV. The central source is clearly
extended up to 3 keV, but examination of the 3-5 keV image shows four
counts detected at the position of the radio core, and no counts in the 5-7
keV band. Contributions from thermal plasma emission should be negligible
in such a small region in the harder bands, so the 3-5 keV counts may arise
from either AGN or X-ray binary emission.

Comparison with the 4.9~GHz contours (derived from VLA archival data,
project AK0360), indicates that the extent of the surface brightness peak
is similar to the distance at which the jets are first detected,
particularly for the western jet. The lack of detectable jets inside this
region could indicate that they have switched off. We consider this
unlikely, since the spectral index in the innermost region where the jets
are detected is $\alpha\sim0.5$, consistent with a currently active radio
galaxy rather than a dying source (GVM08).  We instead assume that the jets
are not detected closer to the core because they are highly collimated and
therefore not resolved. This suggests that the jets lose collimation and
'flare' to a greater width close to the edge of the X--ray surface
brightness feature.

We can estimate the emission likely from unresolved point sources in
NGC~6051 based on known X-ray/optical scaling relations. Unfortunately, the
scale of the surface brightness peak is only a few arcseconds, and no
suitable high-resolution optical data are available. We therefore extract a
$K_s$-band surface brightness profile from the 2-Micron All-Sky Survey
\citep[2MASS,][]{Skrutskieetal06} image of NGC~6051. This image has pixels
1\arcs\ in size, but is derived from a combination of 2\arcs-pixel images
obtained by the survey telescope. The central $K_s$-band surface brightness
profile is effectively smoothed by this process and may be broader and less
peaked than the true profile. 

We estimate the X-ray flux from low mass X-ray binaries (LMXBs) using the
relation of \citet{KimFabbiano04}. The small number of 3-5 keV counts
observed at the position of the radio core are consistent with the expected
LMXB emission in this area, within the (large) uncertainties. We place a
90\% upper limit on the AGN luminosity of L$_{\rm
  AGN}=1.1\times10^{40}$\ergps (0.7-7 keV) by assuming a power law
spectrum with $\Gamma$=1.7, and subtracting our best estimate of the flux
from low-mass X-ray binaries.

\begin{figure}
\centerline{\includegraphics[width=\columnwidth,bbllx=20,bblly=210,bburx=558,bbury=770,clip=y]{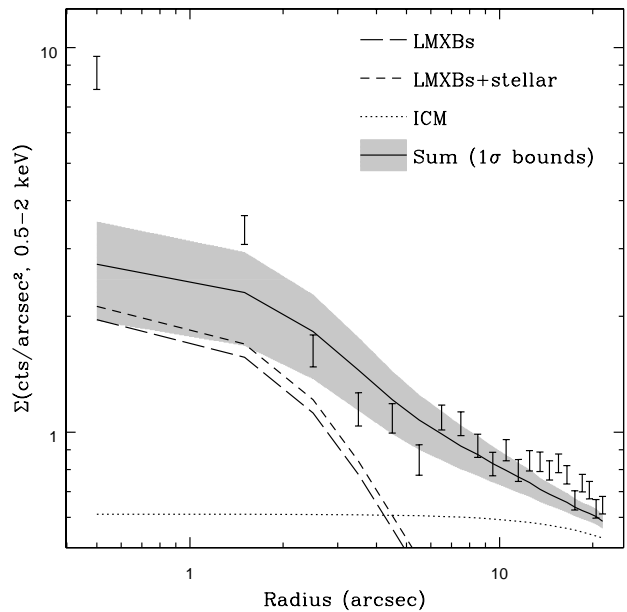}}
\caption{\label{fig:coronasb}0.5-2 keV surface brightness profile of the
  central 20\arcs\ of AWM~4 (errorbars indicate 1$\sigma$ uncertainties),
  compared to estimates of the contribution to surface brightness from
  point sources in NGC~6051 and the intra-cluster medium (ICM). Surface
  brightness profiles for low mass X-ray binaries and for stellar sources
  (cataclysmic variables and coronally active binaries) are estimated by
  scaling the 2MASS $K_s$-band surface brightness profile using the
  relations of \citep{KimFabbiano04} and \citep{Sazonovetal06}.  The ICM
  contribution is determined from a $\beta$-model fit to the emission
  between 30 and 200\arcs. 1$\sigma$ uncertainties on the sum of these
  profiles are estimated from the uncertainties on the scaling relations
  and the normalisation of the surface brightness fit to the ICM.}
\end{figure}

To calculate the expected point source contribution in the area of the
extended soft-band core, we must also include emission from stellar sources
such as cataclysmic variables and coronally active binaries. This
contribution can be predicted using estimates from the Milky Way bulge
\citep{Sazonovetal06}. We work in the 0.5-2 keV band for which this
relation is defined, scaling the LMXB contribution to match. The stellar
sources are assumed to have a 0.5~keV, solar abundance thermal plasma
spectrum. We estimate the contribution from the large-scale gaseous halo of
AWM~4 by fitting a $\beta$-model to the 0.5-2 keV surface brightness
profile between 30\arcs and 200\arcs. This produces an almost flat
contribution to surface brightness in the core. Figure~\ref{fig:coronasb}
shows these predicted surface brightness profiles compared to the measured
surface brightness.

At radii greater than 6\arcs, the sum of the predicted profiles is a
reasonable match to the measured surface brightness profile. Inside this
radius the agreement is less satisfactory; the measured profile is brighter
than expected by a factor $\sim$3 at the centre and appears to be more
sharply peaked. The overestimation of flux at 2-6\arcs, while statistically
acceptable, is probably a result of the poorer resolution of the $K_s$-band
profile. It seems unlikely that the underestimation of the peak flux could
arise from this source, and we therefore conclude that an additional
extended spectrally soft surface brightness component is present.

\section{Spectral Analysis}
\label{sec:spec}

To establish the general radial dependence of gas properties, we extracted
spectra from the S3 CCD in circular annular regions chosen to produce
spectra with a signal-to-noise ratio greater than 100 in a 0.7-7.0~keV
band. A central bin of radius 6 pixels ($\sim$3\arcs) was added to allow us
to search for differences in the central surface brightness peak.  The
annuli were centred on the peak of the X-ray emission, which corresponds to
the optical centre of NGC~6051 and the position of the core of 4C+24.36.
The spectra were fitted using the \textsc{wabs} absorption
\citep{MorrisonMcCammon83} and APEC plasma models \citep{Smithetal01}, and
since gas properties are expected to vary with true three-dimensional
radius, we used the \textsc{xspec projct} model to deproject the data.
Beyond $\sim$200\arcs\ the annuli no longer fall entirely on the S3 chip.
The lost area was corrected for in the \textsc{projct} model, using the
angular coverage parameters for each annulus. The outermost annulus in each
fit contains less than the required number of counts and is not
deprojected; for this reason values in the outermost bin should only be
considered indicative. The radio lobes only extend to $\sim$145\arcs, so we
do not expect edge effects to influence our results.

A second set of annular spectra were also extracted, with the radii of the
bins in the range 3-136\arcs\ chosen to be closely comparable with regions
used in examining the physical properties of the radio jets and lobes.
These spectra are treated identically to the S/N=100 profile, and produce
very similar results. Comparison of the two profiles shows that outlying
data points, such as the temperature in bin 5 of the S/N=100 profile, and
abundance in bin 2 of the alternate profile, are not seen replicated when
the radii of the spectral regions are changed. This suggests that they do
not represent the true temperature or abundance at those radii; the
underlying gas properties are likely to vary smoothly with radius.

\begin{figure}
\centerline{\includegraphics[width=\columnwidth,bbllx=20,bblly=130,bburx=558,bbury=770,clip=y]{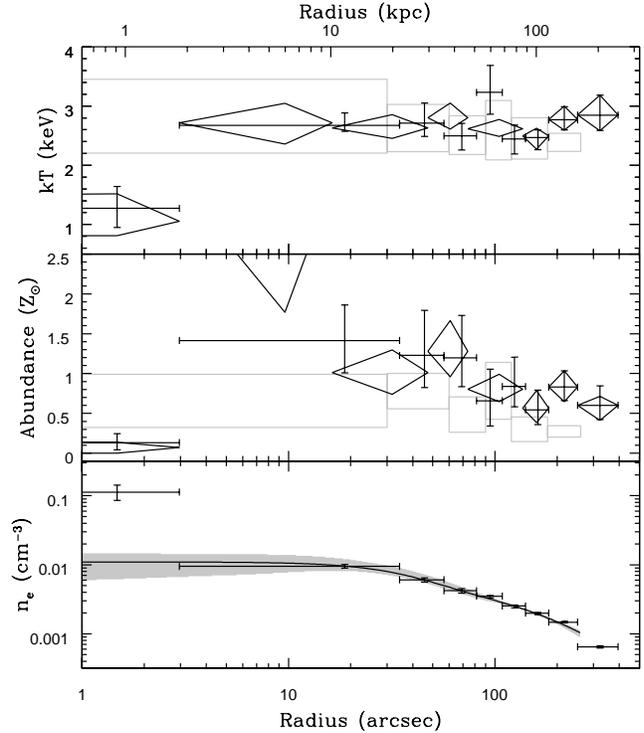}}
\caption{\label{fig:TZ}Deprojected temperature, abundance and density
  profiles for AWM~4. Black diamonds indicate fits to \chandra\ spectra
  from annuli with signal-to-noise ratio $>$100, crosses to a modified set
  of annuli selected for comparison with the radio data. Diamond points are
  omitted from the density plot for clarity, but agree with the crosses
  within uncertainties. Grey rectangles show the \xmms\ deprojected
  temperature and abundance profiles and the solid line shows the density
  profile derived from the \xmms\ data (with 1$\sigma$ uncertainties
  indicated by the grey shaded region), all from OS05.}
\end{figure}

Figure~\ref{fig:TZ} shows the deprojected temperature and abundance
profiles for this deprojection. Comparison with the \xmms\ temperature and
density profiles shows close agreement, confirming the approximate
isothermality and smooth decline of density with radius over the field of
view. The exception is the central bin, which has a temperature of
$\sim$1.2~keV compared to the $\sim$2.6~keV surrounding emission and a mean
density approximately an order of magnitude greater than the ICM. This
suggests that the central surface brightness peak corresponds to a
small-scale cool core or galactic corona.

An apparent difference in abundance profiles between \chandra\ and \xmms\
arises from the different abundance tables used in the analysis. We use the
\citet{GrevesseSauval98} tables, while the older \citet{AndersGrevesse89}
tables were used for the \xmms\ analysis (OS05). Refitting our models
using the older table produces results in good agreement with the \xmms\
profile.  There is some indication of a central abundance plateau in AWM~4,
extending out to $\sim$75\arcs\ ($\sim$50~kpc). Outside this radius the
abundance declines. 

The central bin has a very low abundance, possibly owing to our use of a
single temperature plasma model for a spectrum which must have multiple
emission components. We therefore repeat our deprojection analysis, adding
two more components to the spectral model of the central bin, a 0.5~keV
solar abundance APEC model representing stellar sources and a $\Gamma$=1.7
power law for the LMXBs.  Normalisations were determined from the estimated
surface brightness profiles described in Section~\ref{sec:core}. These
additional components affect the parameters of the original plasma model
only slightly, and we find a temperature of
$k_BT$=1.00$_{-0.17}^{+0.35}$~keV and abundance of
0.10$_{-0.10}^{+0.05}$\Zsol. This is still very low, and it seems likely we
are still underestimating the abundance, possibly owing to the presence of
a range of temperatures within the central bin \citep[The Fe--bias
effect,][]{Buotefabian98}.

\section{Gas Properties of the Core}
\label{sec:coregas}

The mean density in the core, assuming that it fills a sphere of radius
1875~pc ($\sim$3\arcs), equivalent to our spectral extraction region, is
$n_e$=0.101$^{+0.048}_{-0.035}$\pcmcu. \citet{Sunetal07} examined a sample
of small cool cores contained within galaxies, which they argued were
formed largely from material produced by stellar mass loss, distinct from
the surrounding hotter gas associated with the larger group or cluster.
They suggest that these galactic coronae should be in pressure equilibrium
with the surrounding ICM. Estimating the required pressure, we find that a
density of 0.291\pcmcu\ is needed. If the background subtracted flux in the
3\arcs\ aperture arises only from a corona with this density and the
measured temperature, it must occupy a smaller volume, a sphere of radius
1320~pc ($\sim$2\arcs). This is in reasonable agreement with our surface
brightness modelling. The 0.5-2~keV and 1.4~GHz luminosities,
L$_\mathrm{X,0.5-2}=1.76\times10^{40}$\ergps\ and
L$_\mathrm{R,1.4~GHz}=1.31\times10^{24}$~W~Hz$^{-1}$, are in the respective
ranges found for the population of coronae studied by \citep{Sun09}.

The isobaric cooling time of the gas in the cool core can be approximated
as

\begin{equation}
t_{cool} = \frac{5k_BTn_eV\mu_e}{2\mu L_{X,bol}},
\end{equation}

where $V$ is the volume of the gas, $L_{X,bol}$ is the bolometric X-ray
luminosity, $k_BT$ is deprojected temperature, $n_e$ the deprojected electron
number density, and $\mu$ and $\mu_e$ are the mean molecular weight (0.593)
and mean mass per electron (1.167) of the gas. We find a cooling time of
302$^{+180}_{-95}$~Myr, as compared to $t_{cool}$=1.5-3~Gyr for the
$\sim$2.6~keV gas immediately outside the corona.

Gas cooling from the corona may be the main source of material fuelling the
active nucleus. We therefore consider the physical parameters of this
cooling to determine whether the corona is a realistic source of fuel over
the lifespan of the outburst.  The mass of gas cooling out of the X-ray
phase within the corona can be estimated based on the assumption of
continuous isobaric cooling as

\begin{equation}
\dot{M}_{cool} \approx \frac{2\mu m_pL_{X,bol}}{5k_BT},
\end{equation}

where $m_p$ is the proton mass. We find a deposition rate of
$\dot{M}_{cool}$=0.067\Msol yr$^{-1}$. This is a sufficient rate to support
the observed AGN activity; taking our largest estimate of the mechanical
energy output of the AGN, 1.63$\pm0.02\times10^{59}$~erg (for lobes
completely filled by relativistic plasma at the projected distance from the
AGN, see Section~\ref{sec:energy}) and the radiative timescale determined
from the radio observations (170~Myr; GVM08) we can estimate the AGN power
output to be 3.04$\times10^{43}$\ergps. The required efficiency of the AGN
in converting the cooled gas into energy would thus be
$e_{conv}=P_{mech}/\dot{M}c^2$=8$\times10^{-3}$. \citet{Sunetal07} note
that mass loss from stars may contribute to the mass of cool gas available,
reducing the required efficiency further. We estimate the stellar mass loss
from AGB stars within the core to be 0.03\Msol yr$^{-1}$, based on the mass
loss rates of \citet{Atheyetal02}. This suggests that the stellar mass loss
rate may be at least capable of balancing cooling losses from the corona.

If the AGN is fuelled directly from the hot gas of the corona (i.e.,
without the requirement that the gas first cool out of the X--ray regime),
a first approximation of the energy available can be obtained by assuming
the Bondi accretion rate \citep[e.g.,][]{Allenetal06}.  The Bondi accretion
radius $R_A = GM_{BH}/c_a$, where $c_a$ is the adiabatic sound speed
($\sim$536\kmps) and $M_{BH}$ the black hole mass, which we can estimate
from the relationship between $M_{BH}$ and the central stellar velocity
dispersion \citep{Gebhardtetal00}. Taking the averaged velocity dispersion
for NGC~6051 from LEDA\footnote{http://leda.univ-lyon1.fr/},
$\sigma$=343\kmps, we find $M_{BH}$=9.1$\times10^8$\Msol, and therefore a
Bondi accretion radius $R_A$=13.65~pc. The Bondi accretion rate is defined
as $\dot{M}_B = 4\pi R_A^2\rho(R_A)c_a$ where $\rho(R_A)$ is the density at
the Bondi accretion radius. Since this radius is a factor $>100$ smaller
than our innermost spectral bin, it is clear that the uncertainty on
$\rho(R_A)$ will be the dominant source of uncertainty in our estimate of
the accretion rate.

We can place a lower limit on $\rho(R_A)$ by assuming that it is equal to
the mean density of the corona. Unlike cooling time and mass deposition
rate, the Bondi accretion rate will therefore depend on the volume we
assume the corona to occupy. Adopting the lower density derived from the
spectral extraction region we find $\dot{M}_B > 0.002$\Msol yr$^{-1}$,
implying a maximum conversion efficiency of $e_{conv}<0.148$, rather higher
than the efficiency of 0.1 often assumed.  If we instead estimate
$\rho(R_A)$ by fitting two $\beta$-models to the deprojected density
profile, we find a central electron density of $n_e\sim1.2$\pcmcu, about
ten times greater than the mean density of the central spectral bin. This
leads to a Bondi accretion rate $\dot{M}_B=0.022$\Msol yr$^{-1}$ and a
conversion efficiency $e_{conv}=0.014$. This is likely still an
underestimate, as the 0.3-1~keV surface brightness profile, which most
closely reflects the density of the cool gas, does not flatten in the core
at the resolution of \chandra.  However, these values are sufficient to
confirm that accretion directly from the hot phase could fuel the AGN.  We
therefore conclude that at least in principle, the corona can support the
AGN outburst over long periods without the need to build up significant
quantities of cold gas.

Given the strong temperature difference between the corona and ICM,
conduction could have a significant effect on the corona gas.  Studies of
coronae in clusters across a range of temperatures suggest that conduction
must be suppressed (by factors 30-500) for the corona to survive more than
a few Myr \citep{Sunetal07}. This is an important factor in differentiating
coronae from simple cool cores, as the suppression of conduction implies
a physical separation between corona and ICM, with the corona sustained by
stellar mass loss within the galaxy rather than cooling from the cluster.

We assume a temperature gradient of 1.36~keV across a distance of 5.16~kpc,
the distance between the midpoints of the first and second bins of our
temperature profile. We find the mean free path of electrons in the ICM to
be 170~pc. Conductivity follows the relations described by
\citep{Spitzer62} where mean free path $\lambda_e \ll T/|{\rm d}T/{\rm
  d}r|$, the scale height of the temperature gradient, and saturates where
the two are comparable \citep{CowieMcKee77}. We find $\lambda_e\sim185$~pc
and estimate the temperature gradient extends over 5~kpc, so calculate the
``classical'' conduction rate \citep[using the methods described
in][]{OSullivanetal07}.

We find the classical rate of energy transfer to be $\sim$10$^{42}$\ergps,
assuming a radius of 1875~pc for the corona.  The bolometric X--ray
luminosity of the corona gas is only $\sim$5.7$\times$10$^{40}$\ergps, and
radiative cooling therefore cannot balance conduction. The implied
timescale for the corona to be heated to the temperature of the surrounding
IGM is $\sim$10~Myr. If we use the smaller radius estimated from the
pressure balance argument, the conduction rate is reduced, but the corona
will still be heated in $<$20~Myr.  Taking the radiative age of
$\sim$170~Myr as the timescale over which the temperature gradient has
existed, conduction into the corona must be suppressed by at least a factor
15-20, presumably owing to the effects of magnetic fields. Shorter
timescales will reduce the need for suppression, but are unlikely to
resolve the issue.

The estimated mechanical power of the AGN jets would also heat the corona
on short timescales if the jets were interacting with the corona gas.
Assuming a jet power of $\sim10^{43}$\ergps\ (see Section~\ref{sec:energy})
only 0.4 per cent efficiency would be required to heat the corona over the
timescale of the current outburst (170~Myr). The apparent tight collimation
of the jets within the corona suggests that any interaction with the corona
gas is minimal. We conclude that neither conduction nor the AGN jets are
heating the corona to a significant degree. This suggests that the presence
of the corona may disrupt the feedback relationship between gas cooling and
AGN heating, since the jets do not reheat the material which fuels the
nuclear activity.

\section{Properties of the radio source and impact on the ICM}

\subsection{Pressure balance}
\label{sec:PB}
The absence of strong X--ray surface brightness features associated with
the radio source suggests a) that the jets and lobes of 4C+24.36 are in
approximate pressure equilibrium with the ICM, and b) that any cavities
associated with the lobes are only partially filled by radio plasma, or
that some other factor is affecting our estimate of the expected surface
brightness deficit. A comparison of the apparent pressures of the thermal
and relativistic plasma can provide insight into the particle content of
the jets \citep[e.g.,][]{DunnFabian04,Dunnetal05,Birzanetal08}. We
therefore estimate pressure profiles from both the X--ray and radio data.

\begin{figure}
\centerline{\frame{\includegraphics[width=\columnwidth]{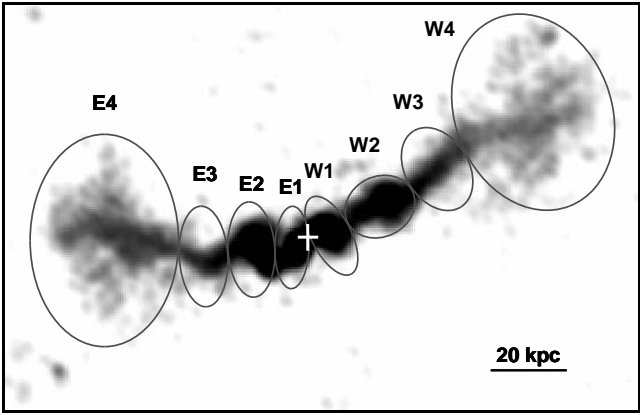}}}
\caption{\label{fig:radioreg} GMRT 610~MHz image of 4C+24.36, with regions
  used to derive radio pressures (selected based on the 235~MHz
    image) marked by ellipses. The white cross marks the position of the
  radio core, at the centre of NGC~6051.}
\end{figure}

Our radio analysis is performed using the GMRT 235, 327 and 610 MHz
observations presented in GVM08 and archival 1.4 and 4.9 GHz VLA data. We
select four regions along each jet, with minimum size based on the lowest
resolution data ($\sim$13\arcs\ HPBW at 235 MHz). These regions are
  shown in Figure~\ref{fig:radioreg}. While they appear large compared to
  the structures in the 610~MHz image, they are well matched to the size of
  jets and lobes at 235~MHz and therefore suitable for examination of the
  spectral index across the different frequencies. The volumes of the
regions are estimated assuming them to be oblate or prolate ellipsoids with
rotation axes aligned with the jet. Fluxes are measured in each band to
obtain the integrated radio spectra for each region. The spectra are fitted
with the Synage++ package \citep{Murgia01}, adopting a continuous injection
model \citep[CI,][]{Kardashev62} to derive the injection spectral index
$\alpha_{inj}$ for each region. Using $\alpha_{inj}$, we then derive the
physical parameters in each regions assuming minimum energy conditions, in
which the contribution to the total energy content from relativistic
particles and magnetic field are approximately equal. Uncertainties on
derived parameters are estimated using the errors on $\alpha_{inj}$
provided by the best-fits, and an assumed 5\% uncertainty on the radio
flux. For our analysis, the synchrotron radio pressure is defined to be

\begin{equation}
{\rm P_{sync}} = U_B + \frac{U_P}{3} = \frac{B_{min}^2}{2\mu_0} +
\frac{(1+k)E_e}{3V\phi},
\label{eqn:prad}
\end{equation}

where $U_B$ and $U_P$ are the energy density of the magnetic field and
relativistic particles, $B_{min}$ and $E_e$ are the minimum energy
magnetic field and total energy in electrons, $\mu_0$ is the permeability
of free space (4$\pi \times 10^{-7}$ N$A^{-2}$ or $4\pi$ if working in
Gauss and {\it cgs} units), $V$ is the volume of the region, $\phi$ is the
filling factor, and $k$ is the ratio of energy in non-radiating particles
to the energy in electrons.

For the initial pressure calculations we assume $\phi=1$ and $k=1$,
implying that half of the energy in particles is in the form of
non-radiating particles, as would be the case in an electron-proton jet.
An electron-positron jet would have $k=0$. We note that our definition of
$k$ differs from some commonly used formulae \citep[e.g.,][]{Fabianetal02}
which use the ratio of the total energy in particles to the energy in
electrons (i.e. $E_P = k_FE_e$). Under our definition, $k=k_F-1$, so our
assumption of equal energy in radiating and non-radiating particles would
imply $k_F=2$.

The pressure of the relativistic plasma also depends strongly on the range
of energies in the particle population. As particles with low $\gamma$
values radiate at frequencies which are not practically observable in most
sources, it is generally necessary to assume an energy range. We have
chosen to consider three possible ranges:

\begin{enumerate}
\item $\gamma$ values equivalent to the frequency range 10~MHz-100~GHz,
  which we refer to as the \textit{standard equipartition} case. This is
  comparable to many previous studies in the literature.
\item $\gamma$=100-5500, referred to as \textit{revised equipartition with
    $\gamma_{min}$=100}. This is the range chosen by GVM08.
\item $\gamma$=10-5500, referred to as \textit{revised equipartition with
    $\gamma_{min}$=10}, chosen to allow us to examine the effect of
  including low energy electrons, under the assumption that the electron
  energy distribution continues to follow a power law down to low Lorentz
  factors.
\end{enumerate}

We discuss these ranges and the reason for their choice below.

Most previous studies of the physical parameters of the jets and lobes of
radio galaxies have adopted the frequency range of 10 MHz-100 GHz normally
used in the standard equipartition equations \citep{Pacholczyk70}. In many
cases high $k$ values \citep[of the order of 100-10000,
e.g.,][]{Dunnetal05} are found, indicating a requirement for a large
fraction of energy in non-radiating particles. Inverse Compton observations
of high-redshift radio sources have shown that the assumption of a power
law distribution of electron energies is reliable down to $\gamma$ values
of a few hundred, which would have synchrotron frequencies $\sim$ 500 kHz.
These lower energy electrons could provide at least part of the additional
pressure implied by the large observed $k$ values \citep{Crostonetal08,
  Dunnetal10}.  GVM08 include these lower energy electrons in their
estimates of total energy and pressure, using the revised equipartition
equations of \citet{Brunettietal97} with a minimum electron energy cutoff
($\gamma_{min}$) instead of a minimum frequency.  Following GVM08, we
estimate the radio pressure assuming $\gamma_{min}=100$ and
$\alpha=\alpha_{inj}$. The maximum energy cutoff, $\gamma_{max}$, was
selected such that the maximum emitting frequency of the electrons matched
the observed break frequency of the radio spectrum, again following GVM08.
In the lobes, this gave $\gamma_{max}$=5500. Approximate break frequencies
for the different regions are given in Table~\ref{tab:breakfreq}.

\begin{table}
\caption{\label{tab:breakfreq} Break frequencies measured in each region of
  the jets and lobes}
\begin{center}
\begin{tabular}{lc|lc}
Region & $f_{break}$ & Region & $f_{break}$ \\
       & (GHz)       &        & (GHz)       \\
\hline
E1 & 30  & W1 & 5   \\
E2 & 3   & W1 & 2   \\
E3 & 1.2 & W3 & 1   \\
E4 & 0.7 & W4 & 0.7 \\
\hline
\end{tabular}
\end{center}
Note that in regions W1 and E1 the frequencies are poorly constrained, as the
change in powerlaw slope is small.
\end{table}

\begin{figure*}
\centerline{\includegraphics[width=\textwidth,bbllx=20,bblly=440,bburx=558,bbury=770,clip=y]{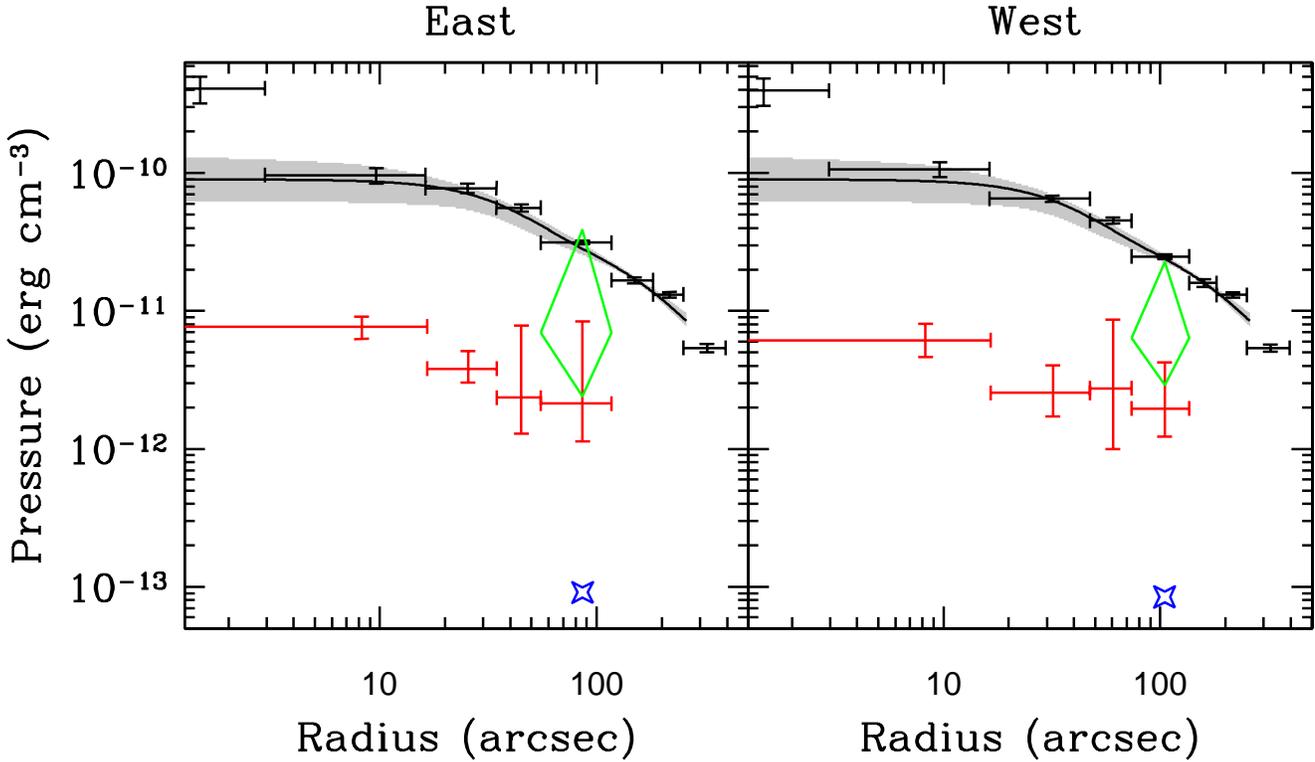}}
\caption{\label{fig:press} Radial profiles of X--ray derived thermal gas
  pressure and equipartition synchrotron pressure for the jets and lobes of
  4C+24.36. Black crosses indicate the thermal pressure derived from
  deprojected \chandra\ spectral models, the black line and grey shaded
  region the \xmms\ pressure profile and 1$\sigma$ uncertainty. The red
  crosses indicate the synchrotron pressure under our assumption of the
  revised equipartition conditions with $\gamma_{min}$=100, and the green
  diamonds the pressure in the lobes if $\gamma_{min}$=10. The blue stars
  represent the ``standard equipartition'' pressure estimates for the
  lobes.  Vertical symbol sizes indicate 1$\sigma$ uncertainties for the
  thermal and revised equipartition synchrotron pressures.}
\end{figure*}

The thermal pressure of the ICM was derived from the deprojected spectral
profiles with annuli selected to match those of the ellipses used in the
radio.  The pressure in each annulus was calculated as P$_{th}=nk_BT$ where
we have assumed an ideal gas with $n=2n_e$. The resulting profile agrees
well with the pressure profile derived from the \xmm\ analysis of OS05,
except in the central bin which the \xmms\ profile does not resolve. We
note that the X--ray pressure profile presented in GVM08 was unfortunately
incorrect, with the normalisation of the thermal gas pressure profile
decreased by a factor of 10 from its true value.  Figure~\ref{fig:press}
shows a comparison between the thermal gas pressure profiles and the
minimum synchrotron pressure estimates described above.

The estimated synchrotron pressure in the jets and lobes is lower than the
thermal pressure at all radii. This is expected; pressure imbalances
between radio sources and the surrounding ICM are common in extended FR-I
radio sources, with the ratio of thermal to synchrotron pressures
(P$_{th}$/P$_{sync}$) having values of up to $\sim$100 \citep[under standard
equipartition assumptions,][]{Ferettietal92}. The uncertainties in the
synchrotron pressures in the outer bins are large. For our
$\gamma_{min}=100$ estimates, the imbalance is $3-5\sigma$ significant,
except in the east lobe where the significance drops to 2.2$\sigma$. While
this means that statistically the east lobe may be in pressure equilibrium,
the consistently low pressure estimates in the other regions strongly
indicate an imbalance in the source as a whole.

The synchrotron pressures decline by a factor 3-4 with radius, as expected,
given radiative energy losses from the particle population of the jets. The
pressure profiles are roughly consistent with a constant pressure
ratio of ${\rm P_{th}/P_{sync}}\sim$15. There is no indication of a strong
gradient in pressure ratio, which might have been expected if the jets are
entraining significant quantities of ICM gas over their whole length.

As our assumed values of $k$=1 and $\phi$=1 appear to be
inconsistent with the data for the $\gamma_{min}=100$ case, we estimate
the change in these parameters which would be necessary to produce pressure
equilibrium in the lobes.  Setting equation~\ref{eqn:prad} to be equal to
the thermal pressure and rearranging allows us to place limits on these
parameters, shown in Table~\ref{tab:kf}.

\begin{table*}
\setlength{\tabcolsep}{1.2mm}
\caption{\label{tab:kf} Magnetic field, pressure, filling factor and
  particle content of the radio lobes (regions E4 and W4)}
\begin{tabular}{lcccccccccc}
\hline
Lobe & ${\rm P_{th}}$ &
  \multicolumn{3}{c}{revised equipartition, $\gamma_{min}=100$} & 
  \multicolumn{3}{c}{standard equipartition} & 
  \multicolumn{3}{c}{revised equipartition, $\gamma_{min}=10$} \\
 & & $B_{min}$ & ${\rm P_{sync}}$ & $(1+k)/\phi$ & $B_{min}$ & ${\rm P_{sync}}$
& $(1+k)/\phi$ & $B_{min}$ & ${\rm P_{sync}}$ & $(1+k)/\phi$ \\
 & (erg~cm$^{-3}$) & ($\mu$G) & (erg~cm$^{-3}$) & & ($\mu$G) &
 (erg~cm$^{-3}$) & & ($\mu$G) & (erg~cm$^{-3}$) & \\[+1mm]
\hline\\[-2.5mm]
East & 31.40$\pm0.98\times10^{-12}$ & 6.36$^{+4.38}_{-1.79}$ &
2.14$^{+6.27}_{-1.00}\times10^{-12}$ & 160.3$^{+243.9}_{-127.3}$ & 
1.92 & 9.11$\times10^{-14}$ & 3192.6 &
11.45 & 6.95$\times10^{-12}$ & 43.3
\\[+1mm] 
West & 24.80$\pm1.00\times10^{-12}$ & 6.09$^{+2.91}_{-2.30}$ &
1.96$^{+2.30}_{-0.73}\times10^{-12}$ & 123.0$^{+135.2}_{-94.2}$ & 
1.85 & 8.44$\times10^{-14}$ & 2448.2 & 
10.97 & 6.38$\times10^{-12}$ & 32.4
\\[+1mm]
\hline
\end{tabular}
Note that values of $(1+k)/\phi$ are those required for pressure
equilibrium in the three cases. The spectral index used was $\alpha_{inj}=1.01$.
\end{table*}

Our alternative estimates of pressure in the lobes can also be considered
in this way. Assuming standard equipartition conditions decreases the
estimated pressure in the lobes by a large factor, with pressure ratios of
$\sim300-350$. This translates into a high value of $(1+k)/\phi$, which
strongly suggests a large amount of additional energy in the non-radiating
particle population. The revised equipartition estimates for
$\gamma_{min}$=10 reduce the pressure imbalance to a factor of 3.9-4.5,
with 1$\sigma$ uncertainties consistent with the thermal pressure. While
these uncertainties are large, owing to the extrapolation of uncertainties
in the spectral index to low $\gamma$ values, this indicates that the
additional energy required for pressure balance could be provided by the
inclusion of electrons with low Lorentz factors. The remaining imbalance
could be explained by relatively minor changes in $k$ or $\phi$; the latter
is of particular interest given the weak cavities in AWM~4. The large
differences between these two estimates clearly indicates the importance of
the equipartition assumptions in determining the pressure imbalance and any
interpretation of the data.

\subsection{Outburst Timescale}

The activity timescale of the radio source can be estimated in several
ways.  Following \citet{Parmaetal86}, GVM08 estimate the age of the source
by modelling the change in radio spectral index along the jets and lobes,
under the assumption that the age of the radio emitting electrons increases
linearly with distance. They find ages of 171$^{+40}_{-15}$~Myr and
159$^{+31}_{-22}$~Myr for the east and west jets and lobes respectively,
assuming $\gamma_{min}$=10 and a fitted injection spectral index
$\alpha_{inj}\sim$0.5. Recalculating with $\gamma_{min}$=100, we find ages
of 183$^{+25}_{-24}$~Myr and 164$^{+24}_{-23}$~Myr for the east and west
jets and lobes. As these calculations are based on multiple independent
spectral index measurements in each direction, this is likely the most
reliable radiative age measurement available. However, it should be noted
that the ages are most applicable to the lobes; a younger age would be
found at a given point along the jet. GVM08 also note that the estimated
age is dependent on the assumption that losses associated with source
expansion are negligible. If this is not that case, the radiative age could
be reduced by up to a factor of $\sim$3.

Radiative ages can also be estimated for individual regions, based on the
minimum energy magnetic field measurements, using Equation~1 of GVM08.  The
injection spectral index estimated for the lobes ($\alpha\simeq1$) differs
from the global estimate ($\alpha_{inj}\sim0.5$), which can lead to large
differences in radiative age.  Adopting $\gamma_{min}$=100, we find an age
of 135$^{+56}_{-66}$~Myr (132$^{+72}_{-52}$~Myr) for the east (west) lobe.
These values are consistent with those derived from the ages derived from
modelling of spectral aging. The standard equipartition assumptions produce
much greater ages, $\sim$252~Myr and $\sim$236~Myr for the east and west
lobes respectively. Conversely, adopting the revised equipartition
assumptions with $\gamma_{min}$=10, we find much shorter ages, $\sim$66~Myr
for both lobes.

Independent limits on the age of the radio source, and therefore on its
magnetic field strength and particle content, can be estimated based on
dynamical arguments. A lower limit on the age of the lobes can be estimated
from the time taken for them to grow to their current size assuming that
(as we see no evidence of shocks associated with them) their expansion has
been subsonic. We can also estimate the time taken for the jets to expand
to their observed length at the sound speed ($\sim$770\kmps\ for
$k_BT$=2.6~keV), though there is a possibility that the jets may have expanded
supersonically earlier in their history and the shocks produced have now
moved out of the field of view or weakened to become undetectable.

A more realistic value may be the time taken for the lobes to rise
buoyantly to their current position. The buoyant velocity is

\begin{equation}
v_{buoy} = \sqrt{\frac{2GM(<R)V}{SR^2C_d}},
\label{eqn:buoy}
\end{equation}

where $R$ is the mean radius from the core at which the lobe is found, $S$
is its cross-sectional area in the direction of motion, $V$ its volume,
$M(<R)$ the total gravitational mass within $R$, and $C_d$ the coefficient
of drag, typically taken to be 0.75 \citep{Churazovetal01}. Finally, we can
also estimate the time required for the ICM to refill the displaced
volume as the lobe rises, $t_{refill} = 2R\sqrt{r/GM(<R)}$, where $r$ is
the mean radial size of the lobe. These four timescale estimates are given
in table~\ref{tab:buoy}.

\begin{table*}
\setlength{\tabcolsep}{1.2mm}
\begin{minipage}{178mm}
\caption{\label{tab:buoy} Radiative and dynamical timescale estimates.}
\begin{tabular}{lccccccccccc}
\hline
Lobe & \multicolumn{2}{c}{$t_{rad,total}$} & \multicolumn{2}{c}{$t_{rad,lobe}$} & $t_{s,lobe}$ & $t_{s,jet}$ & $t_{buoy}$ & $t_{refill}$ & $(1+k)/\phi_{s,jet}$ & $(1+k)/\phi_{buoy}$ & $(1+k)/\phi_{refill}$  \\ 
 & $\gamma_{min}=100$ & $\gamma_{min}=10$ & $\gamma_{min}=100$ &
 $\gamma_{min}=10$ & & & & & & & \\
 & (Myr) & (Myr) & (Myr) & (Myr) & (Myr) & (Myr) & (Myr) & (Myr) &  &  & \\ 
\hline\\[-2.5mm]
East & 183$^{+25}_{-24}$ & 171$^{+40}_{-15}$ & 135$^{+56}_{-66}$ & 66$^{+47}_{-41}$ & 27.53 & 69.5 & 103.16 & 120.02 & 430.3 & 249.6 & 196.3 \\[+1mm] 
West & 164$^{+24}_{-23}$ & 159$^{+31}_{-22}$ & 132$^{+72}_{-52}$ &
66$^{+69}_{-48}$ & 26.77 & 87.3 & 134.76 & 129.20 & 239.0 & 118.8 & 128.7
\\[+1mm]
\hline
\end{tabular}

\medskip
$t_{rad,total}$ is the radiative age estimated from spectral aging along the
jets\\ 
$t_{rad,lobe}$ is the radiative age estimated from the spectral
index of the lobes\\
 $t_{s,lobe}$ is the time required for lobes to expand
to their current width at the sound speed\\
 $t_{s,jet}$ is the time
required for the west jet and lobe to reach their current length at the
sound speed. 
\end{minipage}
\end{table*}

Using these timescales to estimate the magnetic field strength, we can again
assume pressure equilibrium so as to determine $(1+k)/\phi$. However, we must
take into account the dependence of the energy of the particle population
on the magnetic field strength. Following \citet{Brunettietal97}, this
dependence is of the form $E_P\propto B^{1+\alpha}$, leading us to modify
equation~\ref{eqn:prad} as

\begin{equation}
\frac{1+k}{\phi} = \left(P_{th}-\frac{B^2}{2\mu_0}\right)\frac{3V}{C}B^{\,(1+\alpha)},
\end{equation}

where $C$ is a constant related to the synchrotron flux, $\gamma_{min}$ and
the spectral index. We can therefore calculate the change in $(1+k)/\phi$
relative to our original value. The timescales and resulting $(1+k)/\phi$
values, assuming magnetic field strengths derived for the revised
equipartition conditions and $\gamma_{min}$=100, are given in
Table~\ref{tab:buoy}. Estimates based on the sonic expansion timescale of
the lobes are not included as this is short enough to imply synchrotron
pressures greater than the thermal pressure.

The dynamical timescales are all shorter than the GVM08 radiative age
derived from the spectral aging along the jets. The buoyancy and
refill timescales of the west lobe are comparable to the radiative age
derived from the $\gamma_{min}$=100 magnetic field measurement in that
lobe. However, both timescales depend strongly on the filling factor;
reducing $\phi$ would produce a shorter refill time, but a longer buoyant
rise time. The change in buoyancy timescale will depend on the morphology
of the lobe material. For a simple spherical cavity,
$t_{buoy}\propto\phi^{1/6}$, so for $\phi$=0.2, $t_{buoy}$ will increase
by a factor 1.3 to $\sim$175~Myr for the west lobe. The clumpy appearance
of the radio lobes suggests that they have a large surface area for their
volume, so $t_{buoy}$ is likely to increase more slowly. However, it seems
possible that filling factor considerations account for part of the
difference between the buoyancy  and radiative timescales.

The sonic expansion timescale of the jets is comparable to the
$\gamma_{min}$=10 radiative age in the east jet, and longer than it in the
west jet. The large uncertainties on the radiative age mean that there is
no formal disagreement, but this suggests that the jets expanded
supersonically over some portion of their history. If this is the case,
shocks would have been produced, perhaps providing additional heating.

\subsection{Filling factors and Mixing in the Lobes}
\label{sec:ff}
The lack of clearly detected cavities in AWM~4 raises the possibility that
the lobes are not completely filled by relativistic plasma. Some support
for this provided by the radio maps of 4C+24.36 (e.g.,
Figure~\ref{fig:radioreg}), which show both lobes to be clumpy, with
brighter regions apparently tracing the jets out to the tips of the lobes
and fainter, flocculent regions to north and south.  By comparison, the
jets are much more sharply defined.

The lobes may have formed because the jets become uncollimated, with the
relativistic plasma breaking up to form smaller clouds and filaments which
then mix with the ICM. The appearance of the radio emission and the lack of
evidence for higher ICM temperatures along the lines of sight to the lobes
argue against direct mixing of the jet plasma with the ICM gas, instead
suggesting that it remains confined, probably in the form of clumps with
embedded magnetic fields.  For this scenario, we can place limits on the
filling factor $\phi$ based on the X--ray surface brightness in the lobes.

For the western lobe, where only a weak deficit in surface brightness is
seen, we estimate a 3$\sigma$ upper limit of $\phi<0.76$, assuming that
there are no significant additional sources of X--ray flux in the lobe. The
observed decrement of $\sim$40 counts (1120 counts expected, 1080 observed,
1-3~keV) suggests $\phi$=0.21, and is consistent with a zero filling
factor. As discussed in Section~\ref{sec:img}, we would expect a deficit of
184 counts if the lobe were empty of thermal plasma.

In the east lobe, where we see some indication of a cavity, we use the
circular region shown in Figure~\ref{fig:lobes} (as compared to the larger
ellipse discussed in Section~\ref{sec:img}) and find $\sim$430 counts,
giving a decrement of $\sim$60 counts below the $\sim$495 expected from the
surface brightness model.  Assuming ICM properties typical at this radius,
this implies a that only $\sim$24\% of the volume of the lobe is filled by
relativistic plasma. However, the X--ray emission over the remainder of the
lobe is uneven, with weak filamentary structures to north and south of the
holes, so our estimate of the filling factor may be too low. Considering
the lobe region as a whole (i.e., the ellipse region in
Figure~\ref{fig:lobes}), we can place a 3$\sigma$ upper limit of
$\phi<0.43$.

As discussed in Section~\ref{sec:PB}, ${\rm P_{sync}} \propto B^2$ provided
the magnetic field $B$ is similar to the equipartition or minimum energy
field strengths. Using the revised equipartition conditions, the magnetic
field can be defined as

\begin{equation}
B_{eq} = \left(\frac{6\pi(1+k)c_{12}L_{sync}}{V\phi}\right)^{1/(3+\alpha)},
\end{equation}

where $L_{sync}$ is the synchrotron luminosity, $V$ is the volume, and $\alpha$
is the spectral index \citep{GovoniFeretti04}. The constant $c_{12}$ is a
function of the spectral index and the frequency band over which we
integrate \citep{Pacholczyk70}. 

For our source $\alpha\simeq1$, so ${\rm P_{sync}} \propto
\sqrt{(1+k)/\phi}$.  Our filling factors thus suggest a pressure increase
of a factor of 2-2.2, with a 3$\sigma$ minimum increase of $\sim$1.5.  This
is insufficient to resolve the pressure imbalance we estimate from the
revised equipartition assumptions with $\gamma_{min}$=100, but would
significantly reduce the discrepancy. Based on these limits on $\phi$ we
can estimate the value of $k$ (and the 3$\sigma$ upper limits) to be
$k$=37.5 ($k<741.6$) for the east lobe, and $k$=24.8 ($k<517.0$) for the
west lobe.  Filling factors of order $\phi$=0.015 would be required to
match the range of pressure imbalance in the lobes. While this would be
consistent with the lack of surface brightness features in the west lobe,
it is problematic for the decrement seen in the east lobe. This suggests
that simple mixing with the ICM gas cannot account for the observed
pressure imbalance.

If we assume $\gamma_{min}$=10, the limits on $k$ will of course be
reduced. In this case $k$=10.4 for the east and $k$=6.8 for the west lobe.
Within the uncertainties, the $\gamma_{min}$=10 pressures are consistent
with the $k$=1 expected for an electron/proton jet with equal energies in
each class of particles.  However, the uncertainties on $k$ are very large,
and would also be consistent with $k$=0 or with the values estimated using
$\gamma_{min}$=100.

\subsection{Entrainment in the Jets}
A possible additional source of pressure is gas entrained and heated by the
jets. It is thought that entrainment slows FR-I jets and causes their
broadening \citep[e.g.,][]{Worrall09}, and heating of entrained material is
likely to be most effective where the jet is still collimated and its
velocity is high. As we see little sign of interaction between the jets and
corona, the most likely source of entrained material is mass loss from
stars inside the kpc--scale jets. We place a maximum limit on the jet width
from the 0.4\arcs\ resolution of the 4.9~GHz data (250~pc), and take the
length from mean distance between the radio core and the points where the
jets begin to flare (1.5~kpc). We estimate the stellar mass loss rate for
stars within the jets from the mass loss rate of AGB stars in old stellar
populations, $\dot{M}_* = 0.078(L_{\rm B}/10^{10}$\LBsol) \Msol~yr$^{-1}$
\citep{Atheyetal02}, using the mean stellar density within 1.5~kpc,
2.76$\times10^8$\LBsol\ kpc$^{-3}$. Each jet is therefore likely to entrain
approximately 1.63$\times10^{-4}$\Msol~yr$^{-1}$. If we assume the jet has
been active for 170~Myr and is rapid enough to transfer all entrained
material into the lobes, we expect $\sim2.8\times10^4$\Msol\ of stellar
material to be mixed with the relativistic plasma.

Based on the ICM density profile and the ellipsoidal lobe regions assumed
above, we estimate that if they are completely filled by radio plasma, the
lobes have each displaced $\sim3.7\times10^9$\Msol\ of ICM gas. The
pressure imbalance suggests that the radio component in fact occupies only
$<$25 per cent of the lobe volume. For entrained stellar wind gas to
provide the remaining pressure, it would need to be heated to temperatures
$>$5~MeV. Such material could be the source of additional particles in the
jet implied by $k>1$, without contributing significantly to the X--ray flux
from the lobes. It would produce inverse-Compton scattering, but an
electron energy of 5~MeV is equivalent to $\gamma_e\sim10$, so CMB photons
will be scattered into the mid-Infrared rather than the X-ray band.
Assuming that this population provides all of the pressure required to
bring the lobes into equilibrium, and a nominal spread of energies of a
factor of four, we can expect an Infrared flux of
$\sim$1.1$\times10^{-16}$\ergps. This is well below most estimates of the
cosmic Infrared background \citep[e.g.,][]{Doleetal06} and approximately
three orders of magnitude below the 15$\mu$m limit from ISOCAM
\citep{Elbazetal99} which most closely matches the expected energy.

It is also possible that ICM and stellar material are entrained and heated
in the broader section of the jets which extends from the corona out to the
lobes. Since the velocity of the jet is expected to be lower, and the
broader jet is likely to entrain more material, it is possible that the
heating could be mild, producing a separate high temperature thermal
component rather than adding to the jet material. However, if the lobes are
in equilibrium, the maximum energy involved in heating entrained gas is
that required to achieve pressure equilibrium in the volume filled by jet
material. Assuming $\phi$=0.25, this is $\sim5\times10^{56}$~erg, a
significant fraction of the mechanical energy of the jet.

A final consideration is the effect of material entrained within the jet on
the relativistic particle population. Coulomb losses due to collisions with
thermal electrons could be important for particles with low $\gamma$
factors, if the two plasmas are not segregated. \citet{Sarazin99} estimates
the loss rate to be

\begin{equation}
b_{Coul}(\gamma) \approx 1.2\times10^{-12}n_e\left[1.0+\frac{ln
    (\gamma/n_e)}{75}\right] s^{-1}.
\end{equation}

\begin{figure}
\centerline{\includegraphics[width=\columnwidth,bb=20 220 570 750]{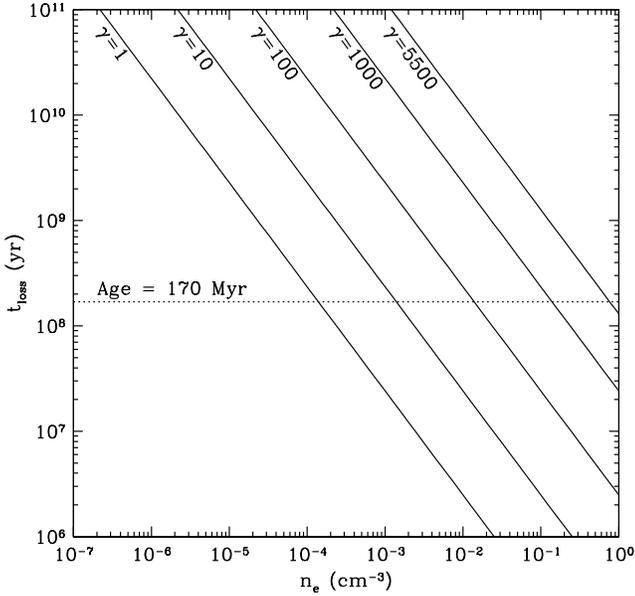}}
\caption{\label{fig:coulomb} The variation in instantaneous loss timescale
  with density, with each solid diagonal line representing particles of a
  given $\gamma$ factor as labelled on the plot. A characteristic age of
  170~Myr is shown for reference.}
\end{figure}

The instantaneous loss timescale $t_{loss}=\gamma/b_{Coul}(\gamma)$ then
gives an approximate timescale over which particles will lose their energy
via Coulomb interactions. Figure~\ref{fig:coulomb} shows the relationship
between loss timescale and density for particles of different $\gamma$
factor.

From the estimated mass of gas lost from stars in the kpc--scale jets, we
can estimate the maximum density of thermal plasma within the jets in this
region. Taking a lower limit on the jet velocity from the large--scale
expansion speed of the radio source (0.002$c$, GVM08), we estimate a
maximum density of $n_e\sim3\times10^{-4}$\pcmcu. This suggests that
Coulomb losses could be significant for particles with $\gamma<10$ but that
more energetic particles will not lose a significant amount of
energy. If stellar mass loss is the only source of entrained material,
losses in the lobes will be even smaller, as the greater mass of gas
occupies a much larger volume.

However, it is clear that if the entrained material is significantly
denser, Coulomb losses will rapidly become important. If densities
$>10^{-3}$\pcmcu occur along much of the jet, we would expect significant
energy losses from particles with $\gamma<100$. In this case the entrained
and heated gas would have to provide the necessary additional pressure,
since the contribution from low--$\gamma$ relativistic particles would be
removed.

We can rule out the most extreme case, entrainment of unheated (or only
mildly heated) ICM plasma by the jets. For the observed ICM densities,
Coulomb losses would affect particles radiating at radio frequencies, and
would produce a clear change in the radio spectrum. This could only be
avoided if the thermal plasma were segregated from the relativistic
particles, with the thermal plasma entrained in non-interacting clumps.

In summary, the lack of clearly detected cavities and the clumpy,
filamentary appearance of the radio lobes suggest that the lobes are
regions of mixing between the relativistic jet plasma and the ICM. Our best
estimate of the filling factor is $\phi$=0.2-0.25. This naturally leads to
a reduction in estimates of the energy in non-radiating particles. If the
energy distribution of the electron population follows the measured
powerlaw to $\gamma=10$, then no additional components are required to
bring the radio plasma into pressure equilibrium with its surroundings,
albeit with large uncertainties. Alternatively, additional material could
be entrained by the jets, probably without producing detectable extra
emission components. However, heating such material to the temperatures
required to produce pressure equilibrium would require a significant
fraction of the total energy of the jets. Coulomb interactions between
entrained material and the relativistic plasma of the jet could also be
important, and while material from stars within the kpc--scale jets is
unlikely to significantly affect the particle population, entrainment of
larger quantities of gas from the ICM could cause the loss of the lowest
$\gamma$ electrons, and any contribution toward the pressure support of the
lobes they provide.

\subsection{Energetics of the radio source}
\label{sec:energy}
To obtain a first approximation of the maximum mechanical energy output of
4C+24.36, we can assume that the radio lobes have in fact excavated
cavities in the ICM, and estimate the power required. The enthalpy in a
cavity is given by $H = \gamma PV/(\gamma - 1)$, where $\gamma$ is the
ratio of specific heats (5/3 for a non-relativistic gas, 4/3 for
relativistic gas). Assuming the cavities to be completely filled with
relativistic synchrotron-emitting plasma, the mechanical energy required is
$4PV$. Assuming the projected distances of the lobes and using the
ellipsoidal regions described in Section~\ref{sec:img} (and shown in
outline in Figure~\ref{fig:lobes}), we estimate this to be
1.63$\pm0.02\times10^{59}$~erg. Taking the age estimate of 170~Myr from
radio spectral modelling, the power output of the AGN would be
3.04$\pm0.04\times10^{43}$\ergps.  The bolometric X--ray luminosity of the
ICM within a sphere with radius equal to the outer radius of the western
lobe is $\sim1.29\times10^{43}$\ergps, neglecting the contribution from the
corona.

At first glance, the similarity between the energy available from the AGN
and the radiative energy loss suggests a balance between heating and
cooling. OS05 discussed the possibility that AWM~4 may at one time have
possessed a large cool core comparable to those observed in clusters of
similar temperature and galaxy population. The estimated that heating such
a cool core to the present typical temperature of 2.6~keV would require
roughly 9$\times10^{58}$~erg. At best only about half of the AGN energy
input would be available to heat the gas after radiative losses, but over
the lifetime of the outburst, the total energy available is
$\sim9.4\times10^{58}$~erg, a close match to energy required.

This simple calculation may underestimate the energy supplied by the AGN.
The timescale of activity could be shorter, implying a greater power output
for the same total energy and smaller fractional losses from radiative
cooling. For a timescale of 135~Myr, similar to the buoyant rise time of
the west lobe, the power of the jets would be $\sim3.8\times10^{43}$\ergps.

As discussed above, there is also the possibility that the jets may have
driven shocks into the surrounding ICM during earlier phases of the
outburst.  Despite the difficulty of observing such shocks, owing to the
short period between their formation and their progression to the low
density outer regions of the cluster halo, several examples have been
confirmed
\citep[e.g.,][]{Gittietal10,Grahametal08,Formanetal07,McNamaraetal05,Nulsenetal05}.
These tend to be weak shocks with Mach numbers $\mathcal{M}\la1.4$, but may
transfer considerable amounts of energy into the ICM. The shocks in the
Perseus and Virgo clusters involve considerably more energy than the radio
lobe cavities in those systems, suggesting that shocks are the dominant
mechanism of energy injection \citep{Formanetal05}, and simulations of
jet/ICM interactions show that multiple shocks may be produced during a
single nuclear activity cycle \citep{Bruggenetal07}. It is therefore
plausible that shocks have contributed to the heating of the ICM in AWM~4
at some stage, though none are currently detected.

However, it seems more likely that our calculation overestimates the energy
input from the AGN. As discussed in Section~\ref{sec:ff}, the lobes are
probably partially filled by relativistic plasma, implying that we have
overestimated the cavity volume. To examine the effect of the filling
factors calculated in section~\ref{sec:ff}, we consider the lower estimate
of filling factor $\phi=0.21$ and the upper limit of $\phi<0.43$.  These
imply enthalpies of 3.4$\times10^{58}$~erg ($<7.0\times10^{58}$~erg) and
mechanical jet powers of 0.64$\times10^{43}$\ergps\
($<1.31\times10^{43}$\ergps), assuming a simple scaling with total volume.
This suggests that the energy available from the radio source is just
sufficient to balance cooling, unless some other mechanism for energy
injection (e.g., shock heating) plays a significant role.

\subsection{Non-thermal X-ray emission from the radio lobes}
\label{sec:nonthermal}
An alternative reason for the lack of clear cavities could be the presence
of non-thermal X--ray emission from the radio lobes, which would ``fill
in'' some or all of the cavity surface brightness decrement. This motivates
us to determine the limits of any inverse-Compton component.

Of the two sources of seed photons available, the energy density of
radio-frequency synchrotron photons produced in the lobes is a factor
$\sim10^{-5}$ lower than that of cosmic microwave background (CMB) photons,
and we therefore neglect synchrotron self--Compton scattering.  We can
estimate the expected X--ray flux from inverse--Compton scattering of CMB
photons from the radio spectral properties, under the assumption of
equipartition.  We can determine which electrons are involved in scattering
from the relation between the change in photon energy and the Lorentz
factor of the electrons, $\nu_{\rm X}/\nu_{\rm CMB} \simeq
(4/3)\gamma_e^2$, where $\nu_{\rm X}$ and $\nu_{\rm CMB}$ are the
frequencies of the scattered X-ray photon and the pre-scattering CMB
photon, and $\gamma_e$ is the Lorentz factor of the scattering electron. To
scatter a photon at the peak frequency of the CMB, $\sim$160~GHz, up to an
energy of 1~keV, an electron with $\gamma_e \sim 1000$ is required
($\gamma_e \sim 330$ for 0.1~keV, $\sim$2800 for 7~keV).  This is within
the range of Lorentz factors used in our equipartition calculations.

\citet{Erlundetal06} provide an estimate of the approximate
inverse--Compton (IC) luminosity in a given energy band, under the
simplifying assumption of a monochromatic distribution of CMB photons:

\begin{equation}
{\rm L_X} = \frac{(4/3)^\alpha}{2} N_0 \sigma_T c \frac{aT_{\rm
    CMB}^4}{\nu_{\rm CMB}^{1-\alpha}}
\frac{\nu_2^{1-\alpha}-\nu_1^{1-\alpha}}{1-\alpha},
\end{equation}

where $\alpha$ is the spectral index determined from the radio data,
$\sigma_T$ is the Thompson scattering cross-section, $c$ is the speed of
light, $aT_{\rm CMB}^4$=$U_{\rm CMB}$, the energy density of the cosmic
microwave background, and $\nu_1$ and $\nu_2$ are the lower and upper frequency
bounds of the chosen energy range. The energy distribution of the electron
population is assumed to be a power law, described by:

\begin{equation}
E = \int^{\gamma_{max}}_{\gamma_{min}} N_0\gamma^{1-p}m_ec^2 \,{\rm d}\gamma,
\end{equation}

where E is the energy of the electron population between Lorentz factors
$\gamma_{max}$ and $\gamma_{min}$, $N_0$ is the normalisation of the
power-law, $m_e$ is the electron mass, and $p$ is the index of the power
law, related to the spectral index of the radio emission and photon index
of the X-ray emission by $p=2\alpha+1=2\Gamma-1$.

Under the assumption of equipartition, the energy of the electron
population in the lobes can be determined from modelling of the radio
spectral distribution. Table~\ref{tab:IC} lists the relevant radio
parameters and resulting flux estimates. We use the injection spectral
index, as this is probably a more accurate estimate of the index at low
values of $\gamma_e$. We note that the energy in the relativistic particle
population, $E_{particles}$, is calculated assuming that electrons and
positrons each contribute 50 per cent of the energy. The expected F$_{\rm
  X}$ is therefore an upper limit; for an electron/proton plasma the
predicted flux will decrease by 50 per cent, and if a larger fraction of
the jet energy is in heavy particles the flux will be further decreased.

\begin{table}
\caption{\label{tab:IC} Expected and measured non-thermal X-ray
  emission in the lobes}
\begin{tabular}{lcc}
\hline
                      & East lobe           & West lobe \\
\hline
$\alpha$              & 1.01                & 1.01 \\[1mm]
$E_{particles}$ (10$^{57}$ erg) & 2.088         & 2.021 \\[1mm]
expected F$_{\rm X}$ (\ergps, 0.7-7 keV)& 1.38$\times10^{-15}$& 1.33$\times10^{-15}$\\[1mm]
measured F$_{\rm X}$ (\ergps, 0.7-7 keV)& $<$8.21$\times10^{-15}$&
$<$4.62$\times10^{-15}$\\
\hline
\end{tabular}
\end{table}

In order to place limits on the actual IC flux in the lobes, we extract
X--ray spectra from each lobe and three annular bins extending from the
lobes out to our limiting radius of 395\arcs, and use a deprojection to
model the contribution of gas along the line of sight to the lobes. The
thermal emission is again modelled with an APEC plasma model, and we add a
powerlaw model with fixed photon index $\gamma=\alpha+1$ to account for the
IC emission. Comparing fits with and without the predicted levels of IC
emission we find no significant differences in spectral parameters. The
expected fluxes are a factor 10$^{-3}$-10$^{-4}$ of the measured total
flux, and the photon indices of $\sim$2 produce a similar spectral shape to
the observed $k_BT$$\sim$2.6~keV plasma spectrum. Allowing the normalisation of
the power-law component to fit freely, we find that the flux is consistent
with the predicted values and with zero, within 90 per cent uncertainties.
The upper bounds on the IC flux are shown in table~\ref{tab:IC}. Even
assuming these maximum IC fluxes to be correct, the parameters of the
plasma model are largely unaffected. We therefore conclude that non-thermal
emission is not detected from the lobes of 4C+24.36, and cannot be
responsible for the lack of distinct X-ray cavities associated with the
lobes.

\subsection{Bending of the jets along the line of sight}
A final factor which could reduce the expected surface brightness decrement
of cavities is bending of the jets. Bending would place the lobes at
greater radii, where the density of the ICM is lower, reducing the pressure
difference.  While GVM08 note that both the small scale jet/counter-jet
brightness ratio and the symmetry of jet properties on large scales suggest
the radio source is aligned close to the plane of the sky, this does not
preclude some bending of the jets along the line of sight.

If we assume that the lobes rise buoyantly and the radiative age estimate
of 170~Myr is correct, we expect the lobe will rise to a distance at which
$M(<R)/R^4=1.73\times10^5$\Msol~kpc$^{-4}$. Using the gravitational mass
profile of OS05, we find that this occurs at $R\sim77$~kpc, compared to
projected distances of $\sim$53~kpc for the east lobe and $\sim$67~kpc for
the west lobe. The thermal pressure at this radius is
$\sim2\times10^{-11}$\ergpcmcu, still a factor $\sim$10 greater than the
synchrotron pressure in the lobes. This would suggest that bending could
contribute $\sim$20\% of the apparent imbalance in the west lobe and
$\sim$35 per cent of the imbalance in the east lobe. Larger contributions
imply higher velocities and pressure balance cannot be achieved within the
radiative timescale as this would require spersonic motion.

If the jets are bent in the line of sight, NGC~6051 must be in motion
relative to the surrounding ICM. A true radius for the lobes of 77~kpc
suggests a galaxy velocity of 316\kmps\ of which 218\kmps\ would be along
the line of sight. However, optical measurements show NGC~6051 to have a
velocity identical (within uncertainties) to the mean velocity of the
cluster galaxy population \citep{KoranyiGeller02}. The uncertainties on
these measurements are small enough that a velocity offset of the magnitude
required appears unlikely. The large difference in luminosity (and hence
mass) between NGC~6051 and the other cluster galaxies, and lack of apparent
substructure in the cluster also argues against recent interactions which
could have produced such a velocity offset. We therefore conclude that
NGC~6051 probably has only a small velocity along the line of sight,
relative to the cluster, and that bending makes only a small contribution
to the apparent pressure imbalance.

\section{Discussion} 
\label{sec:discuss}

The analysis described above provides solutions to a number of the
outstanding questions raised by previous observations of AWM~4, and helps
provide a more coherent context for the cluster among other systems with
powerful central AGN. However, a number of issues remain unresolved.

While there are a number of weak features suggesting that the radio source
4C+24.36 is interacting with the ICM, few are statistically significant
even in this relatively deep \chandra\ pointing. The lack of strong cavities
associated with the radio lobes suggests that relativistic plasma only
partially fills these volumes, mixing with ICM gas which may occupy as much
as 80 per cent of the lobe volume. Radio imaging suggests that the
relativistic component is clumpy and filamentary, suggesting that mixing
has occurred at the level of clouds rather than on microscopic scales.

This evidence of mixing, and the lack of clear cavities raises the question
of why such a process should be observed in AWM~4 but not in other systems.
Any such discussion is necessarily speculative. However, one important
factor is the age of the AGN outburst. We estimate the synchrotron
timescale of the radio lobes to be $\sim$170~Myr, compared to the few tens
of Myr considered typical for FR-I radio galaxies \citep{Macketal98}.
Dynamical age estimates for sources with detected ICM cavities are similar
\citep{Dunnetal05,Birzanetal08}. These estimates may favour younger, more
powerful radio galaxies whose lobes radiate more power at high frequencies
and which can excavate larger cavities in denser environments. However,
even our shortest synchrotron age estimate, 66~Myr, is long in comparison,
and an age of 170~Myr would be fairly unusual. It may be that over such
timescales, energy losses from the relativistic plasma are sufficient for
the confinement of the plasma to weaken, allowing mixing to begin. This
seems most likely to occur if entrainment slows the jet velocity by a large
factor.

The lack of clear cavities is only apparent in AWM~4 because of the
availability of both a deep X--ray observation and low-frequency radio
data. Most studies of cavities have used radio observations at frequencies
$>$1~GHz, and have identified many instances of ghost cavities, X-ray
strcutures with no radio counterpart. Using lower radio frequencies extends
the timescale over which the radio component remains visible, and this may
be allowing us to observe systems which have aged to the point where the
cavities break up. However, deep X--ray observations are also needed, so
that even at large radii, where the expected surface brightness contrast is
small, we can be certain that the cavity is weaker than expected.

There is some evidence of similar ``missing'' cavities in other systems.
One example is HCG~62, in which low-frequency radio observations reveal
emission extending beyond the cavities detected in the X--ray
\citep{Gittietal10}. This suggests the presence of a set of old, outer
radio lobes for which no cavities are detected. There is also the possible
`ancient' bubble in the Perseus cluster \citep{Dunnetal06b} which is
observed only as a temperature structure, with no known radio or X--ray
surface brightness counterpart. The age of this bubble is estimated as
100~Myr, comparable to that of the AWM~4 radio lobes. However, both these
cases differ from AWM~4 in that they appear to be related to old AGN
outbursts, rather than to ongoing activity. Further investigation is needed
to resolve these issues, particularly since it seems likely that an
increasing number of similar cases will be observed as low--frequency radio
observations become more common.

The question of whether mixing occurs only in the lobes or begins through
entrainment of unheated material in the jets remains unresolved. 4C+24.36
has a classic FR-I morphology, with narrow (unresolved) jets in the central
few kpc, broadening rapidly to a width of several kpc over most of their
length, with relatively diffuse lobes and no hot spots. Such jet broadening
is thought to occur because of entrainment of gas within the jets, leading
to a rapid decline of jet velocity \citep[e.g.][]{Worrall09}. Slowing of
the jets may also lead to instabilities which could explain the ``wiggles''
in the jets, as discussed by GVM08. Our estimated values of $k$ could be
explained if external gas has been entrained and heated by the jets. The
large uncertainties mean that we cannot be certain that entrainment is
required, and entrained heated gas is likely not detectable without very
deep hard X--ray observations. Abundance mapping does provide evidence of
gas motions along the jet axis \citep{OSullivanetal10b}, but it is unknown
whether the enriched material is inside the jets or drawn out alongside
them by buoyantly rising lobes \citep{Churazovetal00}.

The jets and lobes of cluster central radio galaxies are commonly found to
be out of pressure equilibrium with the ICM
\citep[e.g.,][]{DunnFabian04,Dunnetal05,Birzanetal08,Crostonetal08}. The
range of pressure differences is extreme, with pressure ratios of one to
several thousand, and the cause of these differences is currently the
subject of debate. This is generally cast in terms of the $k$ parameter, as
higher apparent pressure imbalances imply a need for a greater ratio of
total energy to measured energy in the electron population. 

Examination of cavities and radio bubbles in the Perseus and Centaurus
cluster shows that their $k$ values increase with the radius of the cavity
from the cluster core, and therefore its age \citep{Dunnetal05}, and this
correlation is seen in a small sample of radio sources in galaxy groups
\citep{Crostonetal08}. A correlation is also found between $k$ and the
synchrotron age of cluster radio galaxies, determined from the break
frequency of their radio spectra \citep{Birzanetal08}. These findings can be
interpreted as supporting spectral aging as the cause of the imbalance,
with a larger fraction of the particle population of older sources falling
to lower energies and radiating at lower frequencies. 

Alternatively, entrainment of additional non--radiating articles within the
jets may provide the additional pressure. \citet{Dunnetal06} argue M87 and
NGC~1275 have electron--positron jets on small scales near the AGN, but
lobes with high $k$ values which require the presence of protons (or more
exotic jet models), which must therefore be acquired through entrainment.
\citet{Crostonetal08} show that the apparent pressure imbalance is linked
to morphology, with ``plumed'' FR-I sources (in which much of the lobe is
beyond the end of the collimated jets) have greater imbalances. As
``plumed'' sources have a greater surface area in contact with the ICM,
they interpret the difference in pressure as arising from a greater degree
of entrainment. This result is strengthened by the choice of
\citet{Crostonetal08} to assume a minimum particle energy of
$\gamma_{min}$=10, thereby including an estimate of the additional energy
available from low--energy particles and hopefully reducing the influence
of spectral aging. On the other hand, the characterisation of structures in
the X--ray observations of the systems is hampered by the low X--ray
surface brightness of the systems, and estimation of the filling factor
from the X--ray data is not possible.

Our observations of AWM~4 are well suited to a study of the pressure
imbalance and particle content of the radio lobes. We have high--quality
radio data extending to low frequencies which allow us to place relatively
strong constraints on the synchrotron pressure and age, and make
measurements in multiple regions. We are able to place some constraints on
the filling factors of the lobes, for a source where $\phi\neq1$. Our
source is also old and therefore near--certain to be in equilibrium
with its environment.

It is clear that the choice of energy range over which we assume the
particle population to follow a power law distribution has a dramatic
effect on our estimates of pressure and $k$. Taking the standard range of
10~MHz-100~GHz, we find low synchrotron pressures, high values of $k$, and
therefore a requirement for entrained material. The associated radiative
ages of the lobes are approximately twice the longest dynamical timescales.
It therefore seems likely that the energy of the electron population is
underestimated, and the minimum energy cutoff at 10~MHz is too high.

Using the revised equipartition conditions and $\gamma_{min}$=100 reduces
the pressure imbalance and the required value of $k$ considerably, but does
not reach equilibrium.  Our best estimate of the synchrotron timescale
under these conditions ($\sim$170~Myr, see Table~\ref{tab:buoy}) is rather
longer than the buoyant and refill timescales. However, the low filling
factor of the lobes may bring these timescales into agreement. If we lower
the minimum energy to $\gamma_{min}$=10, pressure equilibrium in the lobes
can be achieved, with $k$ values consistent with minimal (or no)
entrainment of external material.  However, this may imply a radiative age
for the lobes short enough that purely buoyant expansion of the source is
ruled out and supersonic expansion over some part of its lifetime is
required.  This is not physically unreasonable, and may be desirable for
energetic reasons.

We conclude from these results that a large fraction of the apparent
pressure imbalance is due to spectral aging and that inclusion of
low--energy particles by extension to lower values of $\gamma_{min}$
produces more realistic results. We are then left with the question of
whether we consider the shorter timescales associated with
$\gamma_{min}$=10 to be likely.

The relatively flat pressure profile of the jets and lobes (calculated for
$\gamma_{min}$=100) suggests that entrainment from the ICM is not effective
over most of the length of the jet. In this case, if entrained material
does contribute to $k$ it must enter the jet close to its source, on scales
too small for us to examine. This could suggest that entrainment and
heating of external material is only effective while the jet is collimated
and has a high velocity, in its first few kpc in the core of NGC~6051. This
makes supersonic expansion of the jets on scales of tens of kiloparsecs
less likely, and argues for outburst timescales comparable to the buoyant
rise time and higher values of $k$. This is certainly the more conservative
solution. 

We can also consider the thermal state of the ICM. Our estimate of the
total enthalpy of the radio lobes suggests that it is insufficient to
reheat a large cool core comparable to that in MKW~4. However, a smaller
core would require less energy, and additional energy may be available from
shocks or other forms of heating. Using our best estimate of the outburst
timescale, $\sim$170~Myr, the mechanical energy of the jets is comparable
to or slightly less than the rate of energy loss through radiative cooling.
This could indicate that heating was greater at earlier times, or that a
shorter outburst timescale is required. For the $\sim$66~Myr timescale
associated with $\gamma_{min}$=10, the observed filling factors would be
more than sufficient to balance radiative losses from the ICM, and the
supersonically expanding jets would certainly drive shocks.

As discussed at the beginning of this Section, under either scenario the
AGN outburst has a relatively long lifespan. The presence of a corona in
NGC~6051 may provide some explanation. This small volume of cool gas
appears capable of fuelling the AGN outburst, replenished by stellar mass
losses. As the conduction of heat from the surrounding ICM is suppressed,
the corona is likely magnetically separate from its environment. This
implies a breakdown of the ICM--AGN feedback loop, since gas cooling from
the ICM cannot reach the central engine, and the AGN jets do not heat the
corona significantly. If this is the case, the AGN outburst is not
self-limiting and could continue indefinitely.

The origin of the corona is unclear. \citet{Sun09} suggests that powerful
radio AGN may destroy large cool cores in galaxy group scale systems,
leaving only the corona in the central galaxy. The size of the corona may
then be determined by the radius at which the radio jets broaden and begin
to interact with the ICM, or might itself be the cause of this broadening,
with the jets losing collimation as they cross the strong density gradient
(or possibly magnetic field) at its boundary. Associations between jet
broadening and changes in density have been found in other corona-type
systems (e.g., Sun et al. 2005a;b).\nocite{Sunetal05a,Sunetal05b}

In the first case the origin of the magnetic separation between corona and
ICM is unclear unless conduction was suppressed throughout the pre-existing
cool core. Strong suppression of conduction is expected in cool cores owing
to the heat-flux-driven buoyancy instability
\citep[HBI,][]{ParrishQuataert08}. This suggests that where magnetic fields
are aligned radially along the temperature gradient of a cool core,
convection will lead to gas motions which cause the magnetic field to be
realigned tangentially, reducing conduction by large factors. It is unclear
what the timescale for the saturation of the instability would be in a
relatively poor, low-temperature system such as AWM~4 \citep[the timescale
for $\sim$5~keV clusters is estimated to be
$\sim$120~Myr,][]{ParrishQuataert09}, but HBI provides a potentially
feasible mechanism.

The corona could also represent the original galaxy halo of
NGC~6051, compressed and/or stripped by the surrounding cluster gas, with
the galactic magnetic field maintaining separation from the ICM even before
the AGN outburst. The temperature of the material outside the corona prior
to the outburst cannot be known in this scenario.  In both cases, the
separation between corona and ICM suggests that metals released from stars
within the corona cannot easily diffuse outwards, and that any enrichment
must be driven by losses from the stellar population outside the corona.

There is also the possibility that the corona has been stable
over a timescale considerably longer than the current AGN outburst, and
that AWM~4 has been relatively isothermal for a long period. If the corona
is a stable feature, it is difficult to predict the AGN duty cycle, since
ICM cooling would not be the driving force in triggering activity. The
merger of another galaxy with NGC~6051 might disrupt the corona, but as
NGC~6051 is much larger than any other cluster member galaxy and appears
undisturbed, it seems unlikely that a merger has occurred for at least 1
Gyr. For radiative cooling in the ICM to be balanced by energy injection
from the AGN without a feedback mechanism linking the two would be highly
coincidental. As the roughly isothermal temperatures demonstrate that the
AGN is providing at least enough energy to prevent cooling, we would expect
that its energy output over the long--term would exceed radiative losses,
and that no cool core has been able to develop. The relatively steep rise
in gas fraction at $\sim$75~kpc radius reported by OS05 could be an
indication that strong heating has moved ICM gas out of the cluster core.

A final consideration is the motion of NGC~6051 within the cluster. Both
the X-ray morphology and the bending of the radio jets suggest that the
galaxy is moving south relative to the ICM, and the difference in size
between the east and west jets suggests that there may be an eastward
component to the motion. The galaxy population of the cluster is primarily
aligned on a north-south axis \citep{KoranyiGeller02} as is the ICM, so
motion in this direction is perhaps unsurprising. Movement of NGC~6051
relative to the ICM may act to spread the energy released by the AGN more
evenly through the cluster core, and aid in mixing enriched gas outward
from the galaxy.

\section{Conclusions}
\label{sec:con}

We have used a deep, $\sim$75~ks \chandra\ observation of the poor cluster
AWM~4 to examine its structure and properties, and the relationship between
the central radio galaxy and the ICM. Previous studies of AWM~4 found the
cluster to have a number of unusual and conflicting features. GMRT
observations showed that its dominant galaxy hosts an old, active FR-I
radio galaxy, but \xmm\ found no evidence of cooling in the cluster core to
fuel this AGN.  Conversely, heating a cool core to produce the
approximately isothermal $\sim$2.6~keV ICM observed required more energy
than was estimated to be available from the radio source. Our analysis
provides solutions to some of these problems, as well as insights which may
be applicable to other clusters and cluster central radio sources. Our
results can be summarised as follows:

\begin{enumerate}

\item The \chandra\ observation reveals a small cool core located at the
  centre of NGC~6051 and coincident with the core radio source. This meets
  the criteria for a galactic corona \citep{Sunetal07,Sun09}. It is compact
  (radius $\sim$1-2~kpc), significantly cooler than the surrounding cluster
  halo ($k_BT$=1.0$^{+0.35}_{-0.19}$~keV compared to $\sim$2.6 keV for the
  ICM), and has a short cooling time (181$^{+108}_{-57}$~Myr) and moderate
  luminosity (L$_\mathrm{X,0.5-2}=1.76\times10^{40}$\ergps). Heat
  conduction at the Spitzer rate would be sufficient to heat the core to
  the temperature of the surrounding ICM in 10-20~Myr.  This suggests that
  conduction is strongly suppressed. Similarly, a few percent of the
  mechanical energy of the radio jets would be sufficient to have heated
  the corona over the lifetime of the AGN outburst, and we conclude that
  any interaction between the jets and corona must be extremely
  inefficient. VLA 4.9~GHz radio maps do not resolve the jets inside the
  corona, suggesting that they are collimated and narrow, only broadening
  at its outer edge.

\item We estimate the mass deposition rate through radiative cooling of the
  corona gas to be $\dot{M}_{cool}$=0.067\Msol yr$^{-1}$. This would be
  sufficient to power the AGN, requiring an efficiency in converting the
  cool gas to energy of only 0.1 per cent. Mass loss from stars within the
  corona appears sufficient to approximately balance cooling losses from
  the corona.  Direct accretion from the 1~keV gas at the Bondi rate could
  fuel the AGN, though the accretion rate is rather uncertain owing to the
  large extrapolation in radius required.  Magnetic separation of the
  corona from the ICM would prevent gas cooling from the ICM reaching the
  central engine, and the AGN jets do not significantly heat the corona.
  These factors appear to preclude a feedback relationship between AGN and
  ICM. However, the corona itself is capable of fuelling the AGN for long
  periods, and this may explain the unusually long outburst timescale
  estimated for the radio source. It may also explain the lack of a large
  cool core in the system, since without a feedback relationship, AGN
  heating seems likely to have exceeded cooling over the recent history of
  the cluster.

\item Imaging shows the gaseous halo of AWM~4 to be generally relaxed, in
  agreement with previous observations. There are weak indications of
  structures associated with the radio source, including a
  broad bay-like structure around the west jet and lobe. However, the only
  statistically significant surface brightness feature is a decrement near
  the centre of the east lobe. There is no evidence of spectrally hard
  emission associated with the lobes, and the expected level of inverse
  Compton emission is below our detection threshold. If the lobes contained
  only relativistic plasma, we would expect to detect the cavities with
  high statistical significance. We interpret these results as indicating
  that the lobes are only partially filled. This is supported by the
  clumpy, filamentary appearance of the lobes in radio images. Based on the
  surface brightness, we can place limits on the fraction of ICM plasma in
  the lobes. Assuming the remaining volume is occupied by radio--emitting
  relativistic plasma, we find filling factors for this component of
  $\phi=0.24$ and $\phi=0.21$ for the east and west lobes respectively,
  with 3$\sigma$ upper limits of $\phi<0.43$ and $\phi<0.76$.

\item We measure the pressure profile of the ICM, and compare this with
  minimum energy pressure estimates for the jets and lobes of the radio
  source. Under the most conservative conditions, assuming contributions
  from particles emitting between 10~MHz and 100~GHz, we find a strong
  pressure imbalance between the lobes and their environment, with the
  lobes apparently underpressured by a factor $\sim$160. However, these
  conditions imply an age for the source which is considerably longer than
  the timescale for the lobes to buoyantly rise to their current position.
  Estimates which include less energetic particles greatly reduce the
  pressure imbalance, to a factor $\sim$15 for $\gamma_{min}$=100, or to a
  factor $\sim$4 for $\gamma_{min}$=10, with the latter case consistent
  with pressure equilibrium within uncertainties. The radiative age
  estimated for $\gamma_{min}$=100 is roughly consistent with the buoyant
  timescale of the lobes, while the age estimated for $\gamma_{min}$=10 is
  significantly shorter and could imply a supersonic expansion phase for
  the jets. We consider the effects of bending in the jets on our pressure
  estimates, but find that they are unlikely to have a significant impact
  on our results.

\item From these measurements and the filling factor estimates described
  above, we estimate the required ratio of energy in non--radiating
  particles to the observed energy in electrons, $k$, for each lobe.
  Assuming $\gamma_{min}=100$, we estimate $k$=37.5 and 24.8 for the east
  and west lobes, with 3$\sigma$ upper limits of $k<741.6$ and $k<517.0$
  respectively. For $\gamma_{min}=10$, we estimate $k$=10.4 and $k$=6.8 for
  the east and west lobes, with large uncertainties consistent with $k$=1
  (an electron--proton plasma) or $k$=0 (an electron--positron plasma), or
  with the $\gamma_{min}=100$ values. This indicates that in principle the
  apparent pressure imbalance in the lobes can be resolved by the inclusion
  of these lower energy relativistic particles. Alternatively, entrainment
  and heating of thermal plasma (either from stars or the ICM) could
  provide the necessary additional pressure. However, such entrained
  material must have a low density and very high temperature, rendering it
  undetectable with the current data.

\item We estimate the enthalpy of the radio lobes and find that for the
  filling factors estimated above it is $\sim0.3-7.0\times10^{58}$~erg.
  This is lower than the estimated energy required to reheat a large cool
  core such as that seen MKW~4, a cluster of similar temperature and galaxy
  population. The mechanical power output of the jet depends on the
  timescale of the outburst; for our best estimate of $\sim$170~Myr the jet
  power is $\sim0.6-1.3\times10^{43}$\ergps. This is lower than or
  comparable to the bolometric X--ray luminosity of the ICM within the
  radius of the lobes, $\sim1.3\times10^{43}$\ergps, suggesting that in the
  absence of other forms of heating, the energy available from the radio
  lobes is at best just sufficient to balance cooling in the ICM.

  However, the relative isothermality of the ICM and lack of any
  significant cooling region outside the corona strongly suggests that the
  AGN has provided enough energy to balance or exceed cooling losses in the
  past. This could be achieved through additional heating mechanisms (e.g.,
  weak shocks, sound waves, cosmic rays), larger filling factors, or a
  shorter outburst timescale. If we instead assume the shorter radiative
  timescale ($\sim$66~Myr) estimated from the spectra of the lobes and
  assuming $\gamma_{min}$=10, the jet mechanical power is
  2.6$\times10^{43}$\ergps, in excess of the cooling rate. This timescale
  would also require supersonic expansion of the jets, providing additional
  heating through weak shocks.

\end{enumerate}

\noindent{\textbf{ACKNOWLEDGEMENTS}}\\
The authors thank M. Sun and P. Mazzotta for informative discussions, and
the anonymous referee for a number of useful suggestions. Support for this
work was provided by the National Aeronautics and Space Administration
through Chandra Award Number GO8-9127X-R issued by the Chandra X-ray
Observatory Center, which is operated by the Smithsonian Astrophysical
Observatory for and on behalf of the National Aeronautics Space
Administration under contract NAS8-03060. E.  O'Sullivan acknowledges the
support of the European Community under the Marie Curie Research Training
Network. We thank the staff of the GMRT for their help during the
observations. GMRT is run by the National Centre for Radio Astrophysics of
the Tata Institute of fundamental Research. We acknowledge the usage of the
HyperLeda database (http://leda.univ-lyon1.fr).

\bibliographystyle{mn2e}
\bibliography{../paper}

\label{lastpage}

\end{document}